\documentclass{aa}  
\usepackage{graphicx}
\usepackage[caption=false]{subfig}

\usepackage{txfonts}
\usepackage{xcolor}
\usepackage{float}
\usepackage[colorlinks = true,
            linkcolor = blue,
            urlcolor  = blue,
            citecolor = blue,
            anchorcolor = blue,
            breaklinks=true]{hyperref}

\usepackage{blindtext}

\usepackage{booktabs}

\usepackage{multirow}

\usepackage{placeins}

\def\degr{\hbox{$^\circ$}}

\makeatletter
\renewcommand*\aa@pageof{, page \thepage{} of \pageref*{LastPage}}

\begin{document}

   \title{Probing particle acceleration in Abell 2256: from to 16 MHz to gamma rays}


   \author{E. Osinga
          \inst{1,2}
          \and 
          R. J. van Weeren\inst{1}
          \and 
          G. Brunetti\inst{3}
          \and 
          R. Adam\inst{4,5}
          \and 
          K. Rajpurohit\inst{6,7}
          \and
          A. Botteon\inst{3}
          \and 
          J. R. Callingham\inst{1,8}
          \and
          V. Cuciti\inst{7}
          \and
          F. de Gasperin\inst{3,9}
          \and
          G. K. Miley\inst{1}
          \and         
          H. J. A. R\"ottgering\inst{1}
          \and 
          T. W. Shimwell\inst{1,8}
          }

   \institute{Leiden Observatory, Leiden University, PO Box 9513, NL-2300 RA Leiden, The Netherlands 
   \and
    David A. Dunlap Department of Astronomy and Astrophysics, University of Toronto, Toronto, ON M5S 3H4, Canada 
   \email{erik.osinga@utoronto.ca}
   \and
   Istituto Nazionale di Astrofisica, Istituto di Radioastronomia Via P Gobetti 101, 40129 Bologna, Italy 
   \and 
   Laboratoire Leprince-Ringuet (LLR), CNRS, École Polytechnique, Institut Polytechnique de Paris, 91120 Palaiseau, France 
   \and
   Université Côte d’Azur, Observatoire de la Côte d’Azur, CNRS, Laboratoire Lagrange, Nice, France 
   \and 
   Harvard-Smithsonian Center for Astrophysics, 60 Garden Street, Cambridge, MA 02138, USA 
   \and 
   Dipartimento di Fisica e Astronomia, Universitát di Bologna, via P. Gobetti 93/2, 40129, Bologna, Italy 
   \and
   ASTRON, Netherlands Institute for Radio Astronomy, Oude Hoogeveensedijk 4, Dwingeloo, 7991 PD, The Netherlands and Leiden Observatory, Leiden University, PO Box 9513, NL-2300 RA Leiden, The Netherlands 
   \and
   Hamburger Sternwarte, Universit\"at Hamburg, Gojenbergsweg 112, 21029, Hamburg, Germany 
    }

   \date{Received 2023-09-18; accepted 2024-05-07}


  \abstract
    {Merging galaxy clusters often host spectacular diffuse radio synchrotron sources. These sources can be explained by a non-thermal pool of relativistic electrons that are accelerated by shocks and turbulence in the intracluster medium. The origin of the pool and details of the cosmic ray transport and acceleration mechanisms in clusters are still open questions. Due to the often extremely steep spectral indices of diffuse radio emission, it is best studied at low frequencies. However, the lowest frequency window available to ground-based telescopes (10-30 MHz) has remained largely unexplored, as radio frequency interference and calibration problems related to the ionosphere become severe. Here, we present LOFAR observations from 16 to 168 MHz targeting the famous cluster Abell 2256. In the deepest-ever images at decametre wavelengths, we detect and resolve the radio halo, radio shock and various steep spectrum sources. We measure standard single power-law behaviour for the radio halo and radio shock spectra, with spectral indices of $\alpha=-1.56\pm0.02$ from 24 to 1500 MHz and $\alpha=-1.00\pm0.02$ from 24 to 3000 MHz, respectively. {Additionally, we find significant spectral index and curvature fluctuations across the radio halo, indicating an inhomogeneous emitting volume.} In contrast to the straight power-law spectra of the large-scale diffuse sources, the various AGN-related sources that we study often show extreme steepening towards higher frequencies and flattening towards low frequencies. We also discover a new fossil plasma source with a steep spectrum between 23 and 144 MHz, with $\alpha=-1.9\pm 0.1$. Finally, by comparing radio and gamma-ray observations, we rule out purely hadronic models for the radio halo origin in Abell 2256, unless the magnetic field strength in the cluster is exceptionally high, which is unsupportable by energetic arguments and inconsistent with the knowledge of other cluster magnetic fields. 
    } 


   \maketitle
%

\section{Introduction}\label{CH4:introduction}
Galaxy clusters provide a unique laboratory for studying the physics of particle acceleration in cosmic-scale dilute plasmas from the densest and hottest regions of the cosmic web. In these regions, the intracluster medium (ICM) shines brightly both in thermal bremsstrahlung observable with X-ray telescopes \citep{Sarazin1986}, and diffuse synchrotron radio emission due to ultra-relativistic electrons \citep[see][for a recent review]{Weeren2019}. There is significant evidence that both types of emission are driven by the injection of energy through cluster mergers, which heat the ICM and accelerate charged particles through shocks and turbulence \citep[e.g.][]{Markevitch2007,BrunettiJones2014}. 

Because of the dynamic nature of the ICM, galaxy clusters host a panoply of interesting radio sources. Jets of active galactic nuclei (AGN) are found to be more bent closer to the centres of clusters \citep[e.g.][]{Garon2019} or possibly re-accelerated by interactions with the ICM \citep[e.g.][]{Gasperin2017}. On even larger (Mpc) scales, diffuse synchrotron radiation in the form of `radio halos' and `radio shocks' have been widely observed in merging galaxy clusters \citep[e.g.][]{Botteon2022}. In this paper, we adopt the classification of the diffuse synchrotron radiation used in \citet{Weeren2019}, where radio halos are found in the centres of clusters with brightness profiles that generally follow the baryonic distribution of the ICM. In contrast, radio shocks are generally found on the outskirts of clusters and are thought to trace Fermi-I acceleration at shocks \citep{Ensslin1998}. Additionally, there exists another class of diffuse synchrotron sources that are believed to trace old plasma from AGN that has been re-energised by various processes in the ICM. For example, such diffuse emission could have been re-energised by adiabatic compression or internal turbulence. This class encompasses sources such as gently re-energized tails \citep[][]{Gasperin2017} and radio phoenices \citep{Mandal2020}, which can be dubbed `fossil plasma' sources. All classes of diffuse cluster radio emission typically show steep spectra with $\alpha<-1$, where $\alpha$ denotes the spectral index and the radio flux density follows $S_\nu \propto \nu^\alpha$, where $\nu$ denotes the frequency. This implies that the cluster diffuse emission is brighter, and sometimes easier to detect at high significance, at low frequencies.

There are various open questions related to the details of particle acceleration of different classes of diffuse cluster synchrotron sources. One major problem is that the acceleration seen in both the weak radio shocks in the ICM and the turbulent Fermi-II type acceleration in radio halos is not efficient enough to accelerate particles from the thermal pool \citep{BrunettiJones2014}. A `seed' population of mildly relativistic electrons could alleviate this problem both in radio shocks \citep[e.g.][]{Markevitch2005,Kang2012,Botteon2020relics} and radio halos \citep[e.g.][]{Brunetti2001,Cassano2005,Nishiwaki2022}, although the origin of the seed population need not be the same. Possible scenarios for the origin of the seed population are the injection by AGN \citep[e.g.][]{Weeren2017,ZuHone2021b}, multiple weak shocks \citep[e.g.][]{Kang2021}, or secondary products of hadronic proton-proton collisions \citep[e.g.][]{BrunettiBlasi2005,BrunettiLazarian2011,Pinzke2017}. 

The favoured scenario for the origin of radio halos is based on re-acceleration by merger-induced turbulence \citep{Brunetti2001,Petrosian2001,Brunetti2007,Miniati2015,Brunetti2016,Cassano2023}. The role of secondary particles from hadronic interactions in the origin of radio halos is still unclear. A pure hadronic scenario is disfavoured by current radio data and their follow-up \citep[e.g.][]{Brunetti2008,Cassano2010,Bruno2021,Cuciti2021b,DiGennaro2021}. However, the only direct limit to the presence of cosmic ray protons (CRp) and to their contribution to radio halos comes from gamma-ray observations. At the moment, the detection of gamma rays from clusters remains elusive, and the only direct constraints on CRp come from the Coma cluster  \citep{Brunetti2012,Brunetti2017,Xi2018,Adam2021,Baghmanyan2022}.

Abell 2256 is one of the best laboratories for studying particle acceleration mechanisms. This is because of its large angular size and high flux density due to its proximity \citep[$z$=0.058;][]{StrubleRood1999}, coupled with the fact that it is undergoing a massive ($\mathrm{M_\mathrm{500}}=6.2\times10^{14} \mathrm{M_\odot}$; \citealt[][]{PSZ2}) and complex merger. The cluster hosts host clear well-characterised examples of all known classes of diffuse cluster radio emission. It also hosts the lowest redshift radio halo with an ultra-steep spectrum at low frequencies ($\alpha<-1.5$). It has therefore been studied extensively across the electromagnetic spectrum \citep{Briel1991,Briel1994,Rottgering1994,Bridle1976,Bridle1979,Berrington2002,Sun2002,Miller2003,Clarke2006,Brentjens2008,Weeren2009,Kale2010,Owen2014,Trasatti2015,Ge2020,Breuer2020,Rajpurohit2022b,Rajpurohit2023}. 

Until now, neither the ultra-low frequencies ($<100$ MHz) nor the high-energy gamma-rays have been properly explored.  
No good quality data existed on Abell 2256 below 100 MHz due to calibration problems related to the ionosphere, although some ultra-low frequency observations were taken during the early phase of the LOFAR telescope when the calibration and imaging techniques were still in their infancy \citep{Weeren2012Abell}. Those observations, combined with data up to $1.4 GHz$, showed that the radio shock had an unusually flat spectrum of $\alpha=-0.81\pm0.03$, inconsistent with standard diffusive shock acceleration and that the radio halo showed unexpected flattening towards higher frequencies. These results were however not corroborated by recent higher frequency investigations \citep{Rajpurohit2022b,Rajpurohit2023}. A more thorough ultra-low frequency study of Abell 2256 is therefore warranted to accurately quantify and characterise the low-frequency emission. 

Recent advances in calibration and imaging techniques have made routine LOFAR Low Band Antenna (LBA) observations at $\sim 50$ MHz possible \citep[e.g.][]{Gasperin2019,Gasperin2020,Biava2021,Edler2022,Botteon2022SciA,Gasperin2023}. In principle, the LOFAR LBA system works down to 10 MHz \citep{Haarlem2013}, but no standard data reduction pipeline yet exists for observations in the 10-30 MHz range. 

In this paper, we present the deepest radio images made at the lowest radio window available to ground-based telescopes. We study particle acceleration in Abell 2256 by combining those data with higher-frequency data from the literature and gamma-ray upper limits from 13.5 years of Fermi-LAT observations. A flat concordance cosmology with $H_0=70$ kms$^{-1}$Mpc$^{-1}$, $\Omega_m$=0.3 and $\Omega_\Lambda=0.7$ is adapted, which means that at the cluster redshift, 1 arcsecond corresponds to 1.12 kpc.


\begin{figure*}[thb]
    \centering
    \includegraphics[width=1.0\textwidth]{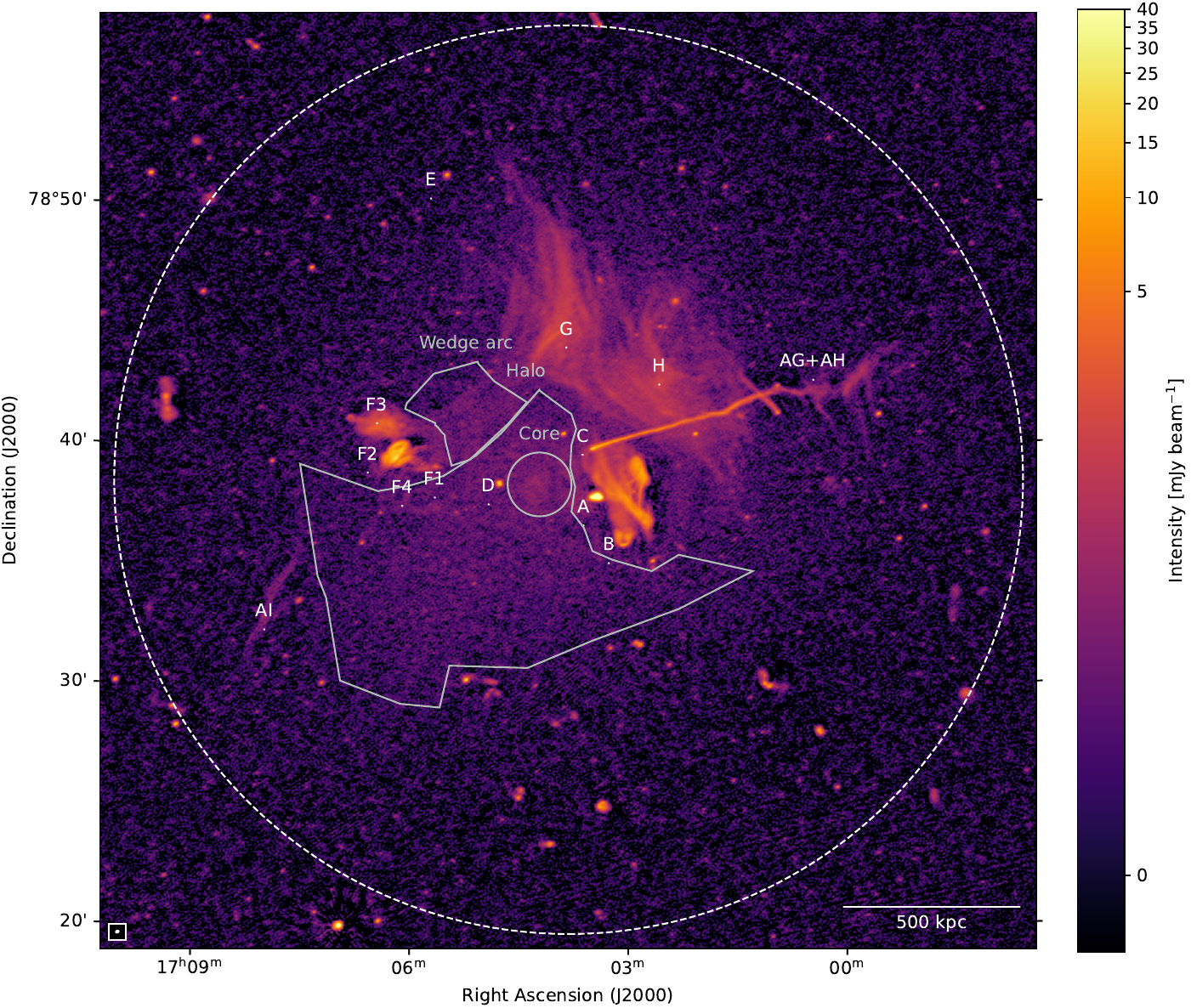}
    \caption{Full resolution LOFAR HBA (144 MHz) image of Abell 2256 with the $6^{\prime\prime}\times6^{\prime\prime}$ HPBW restoring beam, shown in the bottom left corner. The background rms noise is 90 $\mu$Jy beam$^{-1}$ and sources are labelled according to \citet{Rottgering1994}. The dashed circle indicates the $R_\mathrm{500}$ radius.}
    \label{fig:radiobig}
\end{figure*}

\begin{figure*}[thb]
    \centering
    \includegraphics[width=1.0\textwidth]{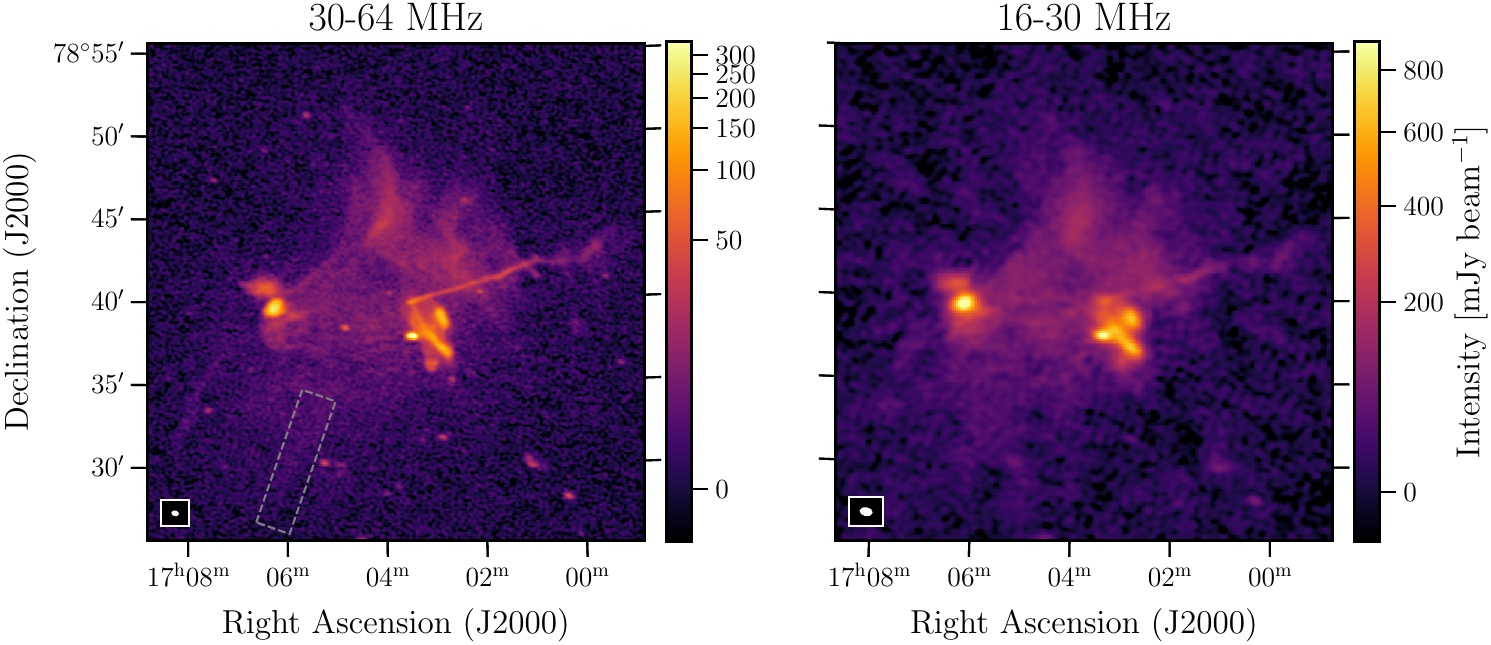}
    \caption{Full resolution LOFAR LBA (24 and 47 MHz) images of Abell 2256 at different frequencies with the restoring beam shown in the bottom left inset. The 46 and 23 MHz images, respectively, have resolutions (HPBW) of 19$^{\prime\prime}\times12^{\prime\prime}$ and 39$^{\prime\prime}\times24^{\prime\prime}$, and background noise levels of 1.4 mJy beam$^{-1}$ and 9 mJy beam$^{-1}$. The tentative filamentary extension of the halo, parallel to source AI, is indicated by the grey dashed region.}
    \label{fig:radiofullres}
\end{figure*}

\begin{table*}[tbh]
\centering
\caption{LOFAR radio observations used in this work.}
\label{tab:radioobs}
\begin{tabular}{@{}lllll@{}}
\toprule
Start date\tablefootmark{a} [UT] & Frequency [MHz] & Integration time [hours] & Distance to pointing  centre\tablefootmark{b} [deg] & LOFAR Project code \\ \midrule
2018-05-01 21:11 & 120-168 & 8 & 1.6 & LC9\_008 \\
2019-05-31 19:20 & 120-168 & 8 & 1.0 & LT10\_010 \\
2021-01-08 09:00 & 16-64 & 4 & 0.0 & LC15\_026 \\
2021-01-10 08:00 & 16-64 & 8 & 0.0 & LC15\_026 \\
2021-01-16 09:00 & 16-64 & 4 & 0.0 & LC15\_026 \\ \bottomrule
\end{tabular}
\tablefoot{\tablefoottext{a}{Dates are given in the yy-mm-dd format}\tablefoottext{b}{The distance from the cluster Abell 2256 to the pointing centre of the observations.}}
\end{table*}

\section{Data}\label{CH4:data}

The radio observations used in this work are listed in Table \ref{tab:radioobs}. Abell 2256 was observed with both the LOFAR LBA and High Band Antenna (HBA) systems for 16 hours each. The observations and calibration process are detailed below, separately for the HBA and LBA. The flux density scale of both systems was verified with bright compact sources in the field using recent higher frequency data from \citet{Rajpurohit2022b}, as shown in Appendix \ref{appendixI}. We found that the LOFAR HBA maps were biased slightly high, and thus corrected those with a scaling factor of 0.83 \citep[see also][]{Rajpurohit2022b}. Throughout the paper, a 10\% uncertainty will be assumed on all flux measurements, which is common for LOFAR observations \citep{Shimwell2022}. All images are made using Briggs weighting with a value of the robust parameter equal to -0.5 \citep{Briggs1995}.

\subsection{LOFAR HBA}
The LOFAR HBA (120-168 MHz) observations were taken in the $\mathtt{DUAL\_INNER}$ configuration (i.e. the remote station collecting area is matched to the core stations) in two different observing sessions. Observations taken on 2018-05-01 include Abell 2256 at a distance of 1.6 degrees from the pointing centre as part of the LOFAR project with code \textit{LC9\_008}. Additionally, observations from the LOFAR Two-Metre sky survey \citep{Shimwell2017,Shimwell2019,Shimwell2022} of the field P255+78, include Abell 2256 at a distance of 1.0 degree from the pointing centre. 
The total target observation time is 16 hours spread equally over the two observations and both observations were book-ended with 10-minute scans on the calibrator source 3C295. 

We separately calibrated both observations using the standard LoTSS DR2 pipeline \citep[full details in][]{Tasse2021,Shimwell2022}. First, direction-independent effects such as polarisation alignment, Faraday rotation, bandpass and delay terms were corrected in $\mathtt{prefactor}$\footnote{\url{https://git.astron.nl/eosc/prefactor3-cwl}} using the calibrator observations \citep{Weeren2016,Williams2016,Gasperin2019}. The solutions were applied to the target field after which several cycles of direction-dependent (self-) calibration were done.

After the complete direction-dependent calibrated image was created with the standard LoTSS DR2 pipeline, we extracted a region of $0.5\degr\times0.5\degr$ around Abell 2256, using the extraction procedure detailed in \citet{Weeren2021}. This optimises the image quality of the main target of interest by removing sources away from the target and performing a direction-independent self-calibration towards the target with the full 16-hour dataset. The resulting image is shown in Figure\,\ref{fig:radiobig}. The final science image has a background RMS noise of 90 $\mu$Jy beam$^{-1}$ when imaged at the half-power beamwidth (HPBW) resolution of $6^{\prime\prime}\times6^{\prime\prime}$.

\subsection{LOFAR LBA}
Abell 2256 was observed with the LBA system as part of LOFAR project \textit{LC15\_026} from 16 to 64 MHz in three separate observing runs, detailed in Table \ref{tab:radioobs}. We employed a similar observing strategy to the LOFAR LBA sky survey \citep[LoLSS; ][]{Gasperin2021}, observing a calibrator source (3C380) simultaneously during the entire run. Similar to the LoTSS DR2 pipeline and LoLSS pipelines, we first used $\mathtt{prefactor}$ over the full bandwidth to calculate the direction-independent corrections, which may vary over the time of the observation. These corrections include the polarisation alignment, bandpass and LOFAR beam model. 
Afterwards, calibration was performed separately for the frequency range 16-30 MHz and 30-64 MHz. 

\subsubsection{30-64 MHz}
For the frequency range 30-64 MHz, we used the pipeline employed for LoLSS\footnote{\url{https://github.com/revoltek/LiLF}} \citep{Gasperin2019,Gasperin2020}. This pipeline first solves for direction-independent effects \citep{Gasperin2019} in the target field by self-calibration, starting from a model from TGSS ADR1 \citep{Intema2017} and then direction-dependent effects as described in \citet{Gasperin2020}. After successful calibration and imaging of the complete field-of-view, we extracted the target cluster using the method detailed in \citet{Weeren2021}. The final image integrated from 30-64 MHz has a resolution of 19$^{\prime\prime}\times12^{\prime\prime}$ and an rms noise of 1.4 mJy beam$^{-1}$. It is shown in the left panel of Figure\,\ref{fig:radiofullres}.

\subsubsection{16-30 MHz}
For the lower part of the LBA sub-band, from 16 to 30 MHz, no standard pipeline is yet available, although \citet{Groeneveld2024} recently presented a calibration strategy for the decametre band that is shown to work for a standard LOFAR observation of an arbitrary field with typical observing conditions. We have used a similar method to calibrate the Abell 2256 field, proceeding as follows.
We re-calculated phase calibration solutions in two steps using the calibrator source and solution intervals and smoothness constraints optimized for the frequency range. First, differential Faraday rotation was calibrated by converting the data to a circular basis and taking only the phase difference of the XX and YY correlations. This has the advantage that all scalar phase effects are removed from the data.
Then, scalar phase effects (i.e. ionospheric dispersive delay and clock terms) were taken out by solving for a model of the calibrator source. For both of these calibrations, we constrained the solutions to be smooth by convolving them with a Gaussian kernel that has a width that is linearly proportional with the frequency, to follow the $\nu^{-1}$ dependence of ionospheric dispersive delays.
The calibrator phase solutions were then applied to the target field, which concluded the data pre-processing. The first direction-independent image was then made by means of self-calibration using a bright calibrator in the target field, that dominates the flux density. We phase shifted to the brightest source in the target field, 3C390.3, and used the same calibration strategy as for the calibrator field, which solves for differential Faraday rotation and residual phase effects, but now in the direction of the target field. We used again the TGSS-ADR1 survey as the starting model. 
Finally, for direction-dependent calibration of the target field, we manually extracted $\sim 1^\circ \times 1^\circ$ regions around the 13 brightest sources in the field. Those were self-calibrated to correct for ionospheric distortions by calibrating for total electron content (TEC) and phase simultaneously ($\mathtt{tecandphase}$ in DP3; \citealt{vanDiepen2018}), again using the TGSS-ADR1 survey as a starting model. The final direction-dependent calibrated image was made by combining the solutions from different directions to a smooth screen. 

The full field-of-view of the LBA image is shown in the appendix (Fig. \ref{fig:fullFOVLBA}), where the imaging was done in WSclean using multi-scale clean \citep{Offringa2014,Offringa2017} and the image-domain gridder \citep{vanderTol2018}. Then, as was similarly done for the higher frequency data, we manually extracted the direction of the target and performed additional rounds of self-calibration to optimise the calibration quality in the direction of Abell 2256 \citep{Weeren2021}. The right panel of Figure\,\ref{fig:radiofullres} shows the resulting image of Abell 2256. We achieved unprecedentedly low Gaussian noise levels ($<$ 10 mJy beam$^{-1}$) in the frequency range 16-30 MHz. This presents not only the deepest-ever image of Abell 2256 at such low frequencies but also of any celestial target. 

\subsection{Gamma-ray data}
For comparison with gamma-ray observations, we have made use of publicly available data from the Fermi Large Area Telescope. The event selection and analysis follow the work presented in \cite{Adam2021}. 

We used 13.5 years of Pass 8 data (P8R3), collected from August 4, 2008, to February 7, 2022. They were extracted within a radius of 10 degrees from the cluster centre. We selected events with energies from 200 MeV to 300 GeV and we applied the $\mathtt{P8R3\_SOURCE\_V2}$ selection (event class 128) and selected $\mathtt{FRONT+BACK}$ converting photons (event type 3). Data from zenith angles less than 90 degrees were filtered out to remove the Earth limb photons. Time selection and rocking angle cuts were applied following recommendation: $\mathtt{DATA\_QUAL}>0$ \&\& $\mathtt{LAT\_CONFIG==1}$, and $\mathtt{(ABS(ROCK\_ANGLE)<52)}$.

Here, we focus mainly on the gamma-ray spectral constraints, to be combined with radio synchrotron data. In order to extract the cluster SED, we performed a joint likelihood fit of both the background components and the cluster using the \textit{fermipy} package \citep{Wood2017}. The data were binned both in energy and space, with 8 energy bins per decade and 0.1x0.1 deg$^2$ pixels. The region of interest (ROI) width was set to 12 degrees. We model the ROI using the 4FGL-DR2 catalog ($\mathtt{gll\_psc\_v20.fit}$; \citealt{Abdollahi2020, Ballet2020}) together with the isotropic diffuse background ($\mathtt{iso\_P8R3\_SOURCE\_V2\_v1.txt)}$ and the galactic interstellar emission ($\mathtt{gll\_iem\_v07.fits}$). The cluster gamma-ray template was modelled using the MINOT package \citep{Adam2020}. MINOT requires a thermal gas model and a cosmic ray proton (CRp) spatial and spectral distribution to compute gamma-ray templates from hadronic interactions. {The thermal model was fixed to the one discussed in Section \ref{CH4:Gammaray}}. When fitting the sky model to extract the SED, the photon spectral index is allowed to vary within the bins so that the final results are insensitive to the CRp spectrum. Given the fact that Abell 2256 is barely resolved by the Fermi-LAT, the SED constraints are only weakly sensitive to the assumptions made about the CRp spatial distribution with the cluster (see Section \ref{CH4:Gammaray} for the modelling). We performed the spectral extraction using different assumptions about the spatial modelling and concluded that the results remained stable. In the end, we obtained, in each energy bin, the likelihood scan for the normalization of the flux that is either used to constrain the cluster CRp normalisation and spectrum independently from other wavelengths, or used jointly with radio data for testing acceleration models (Section \ref{CH4:Gammaray}).

\section{Results - Radio analysis}\label{CH4:results}

\begin{figure*}[thb]
    \centering
    \includegraphics[width=1.0\textwidth]{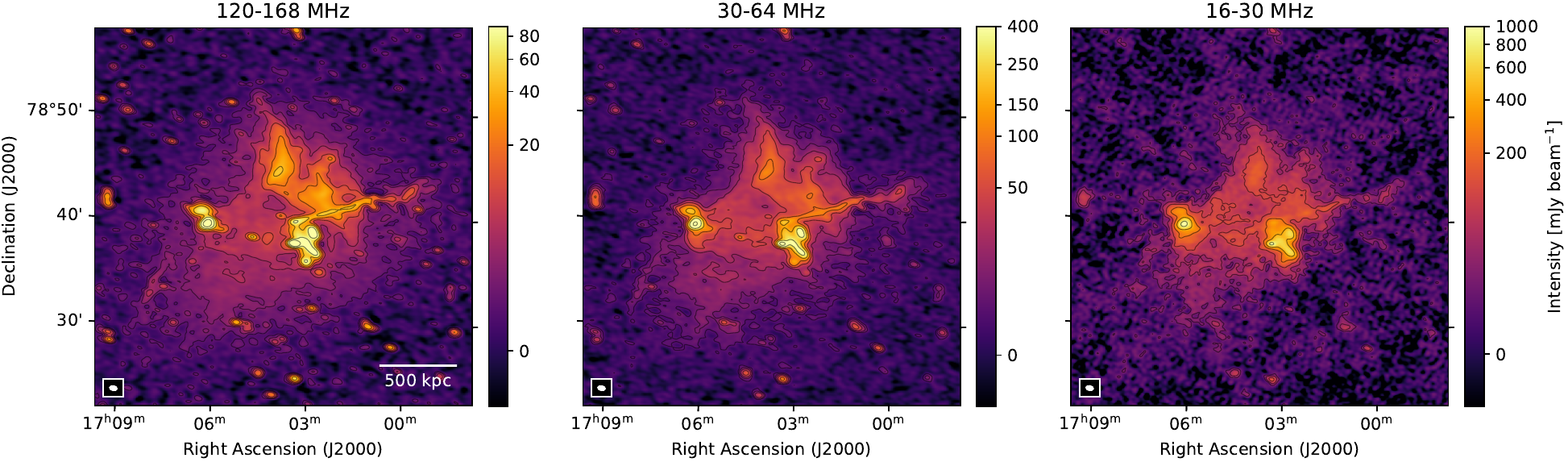}
    \caption{Common resolution (39$^{\prime\prime}\times24^{\prime\prime}$) LOFAR images of Abell 2256 at different frequencies with the restoring beam shown in the bottom left inset. The background noise levels are $\sigma=$0.4, 1.7 and 9 mJy beam$^{-1}$ respectively. Contours are drawn at [3,6,12,\dots,384]$\sigma$.}
    \label{fig:radiolowres}
\end{figure*}

\begin{figure*}[thb]
    \centering
    \includegraphics[width=1.0\textwidth]{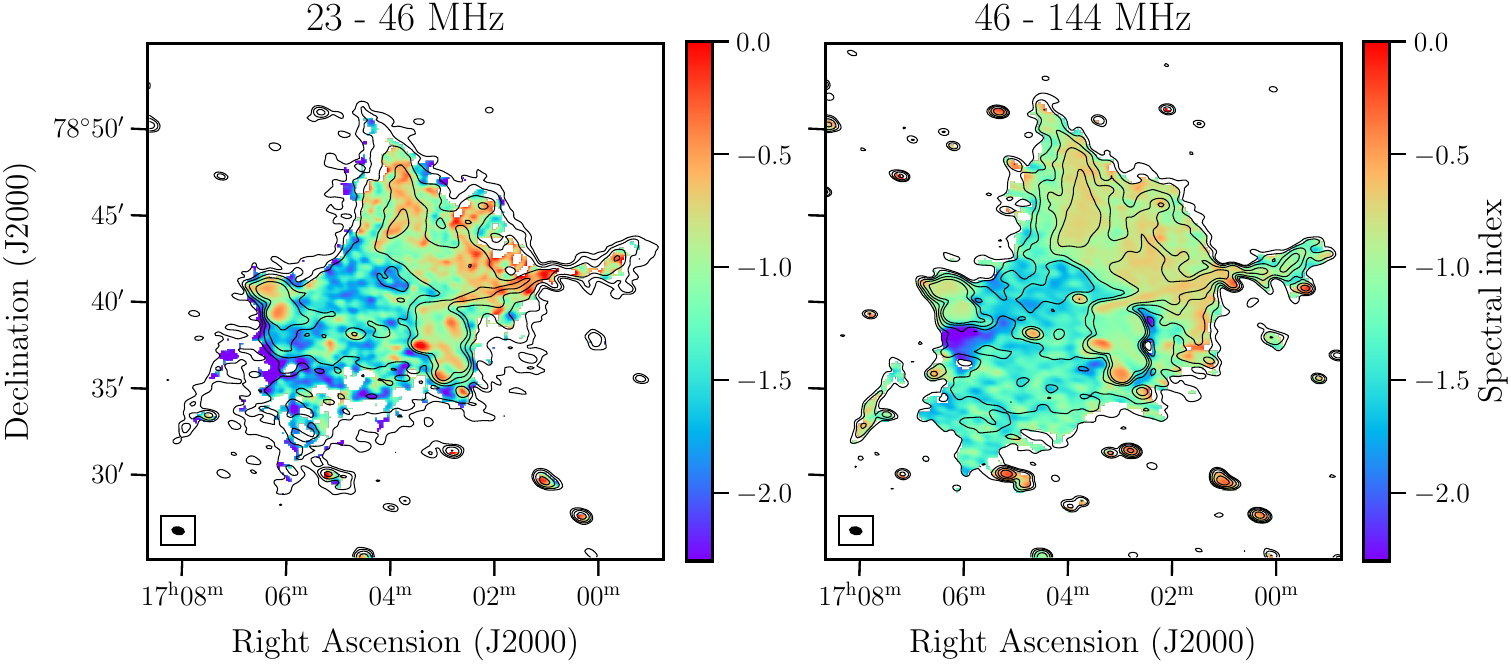}
    \caption{Spectral index maps of Abell 2256 at different frequencies with the restoring beam shown in the bottom left inset. Both maps have been smoothed to a common resolution of 39$^{\prime\prime}\times24^{\prime\prime}$ and were made with an inner \textit{uv} cut of 100$\lambda$. The contours show the higher frequency [3, 6, 12, 24, 48]$\sigma$ levels where $\sigma$ denotes the background rms noise level.}
    \label{fig:radiofullspix}
\end{figure*}

The full-resolution LOFAR images are shown in Figures \ref{fig:radiobig} and \ref{fig:radiofullres}, where no \textit{uv} baseline filtering is applied. Although the resolution is low in the 16-30 MHz band, we can still clearly distinguish the distinct sources of radio emission in the cluster. The radio halo and radio shock are clearly resolved, and the brightest AGN-related sources can still be separated from the diffuse radio emission. 

To emphasise the low surface brightness emission in the cluster, we plot the three frequency bands all convolved to the same resolution of 39$^{\prime\prime}\times24^{\prime\prime}$ in Figure\,\ref{fig:radiolowres}. We note that at the matched resolution, the 30-64 MHz and 16-30 MHz images have a similar sensitivity to the HBA system for sources with a spectral index of $\alpha=-1.3$ and $\alpha=-1.7$, implying that the HBA image is more sensitive than the two LBA images for sources that have $\alpha>-1.3$ and $\alpha>-1.7$, respectively. 

We also made two spectral index maps, between 16-46 and 46-144 MHz, at a common resolution of 39$^{\prime\prime}\times24^{\prime\prime}$.
For these maps, we set the robust parameter to -0.5 and employed an inner \textit{uv} baseline cut at 100 times the observing wavelength (i.e. $100\lambda$), to ensure short baselines are similarly sampled at all frequencies. Additionally, only pixels with a flux density greater than three times the RMS noise in all three images were used. Figure\,\ref{fig:radiofullspix} shows the spectral index maps, with contours representing the total intensity of the higher-frequency image. The spatial distributions of uncertainties are shown in the Appendix Figure \ref{fig:spix_err}, {calculated from Eq. \ref{eq:spixunc}} 
including both the flux scale offset and statistical uncertainty. The median spectral index uncertainty is $0.31$ for the lower part of the LBA band and $0.19$ for the LBA-HBA map. In the following sections, we present the analysis of the radio halo, radio shock and AGN-related sources separately.

\subsection{Radio halo}\label{CH4:radiohalo}
In the low-resolution images shown in Figure\,\ref{fig:radiolowres}, the halo appears largest at 144 MHz, owing to the high sensitivity of the HBA system. In fact, with the low-resolution image bringing out the low surface-brightness emission, the radio halo is larger than reported in the recent work by \citet{Rajpurohit2023}, with a largest-linear size (LLS) of 0.40$^\circ$, corresponding to 1.6 Mpc at the cluster redshift. This is due to the fact that the halo LLS was measured in $20^{\prime\prime}$ images with a \textit{uv} cut of $100\lambda$ by \citet{Rajpurohit2023}, but these full uv plane lower resolution images show that low surface-brightness emission extends further. 
The halo encompasses the radio shock, extending out to about 60\% of the $R_\mathrm{500}=1273$ kpc \citep[][]{PSZ2} radius. 
The halo emission appears to become filamentary in the south-east region, which is best visible in the high-resolution 120-168 and 30-64 MHz images in Figures \ref{fig:radiobig} and \ref{fig:radiofullres}, where the latter has been marked to indicate the location of the possible filament. We note that the filamentary emission is oriented approximately parallel to the source AI, but also underline that their detection remains tentative.

\begin{figure}[thb]
    \centering
    \includegraphics[width=1.0\columnwidth]{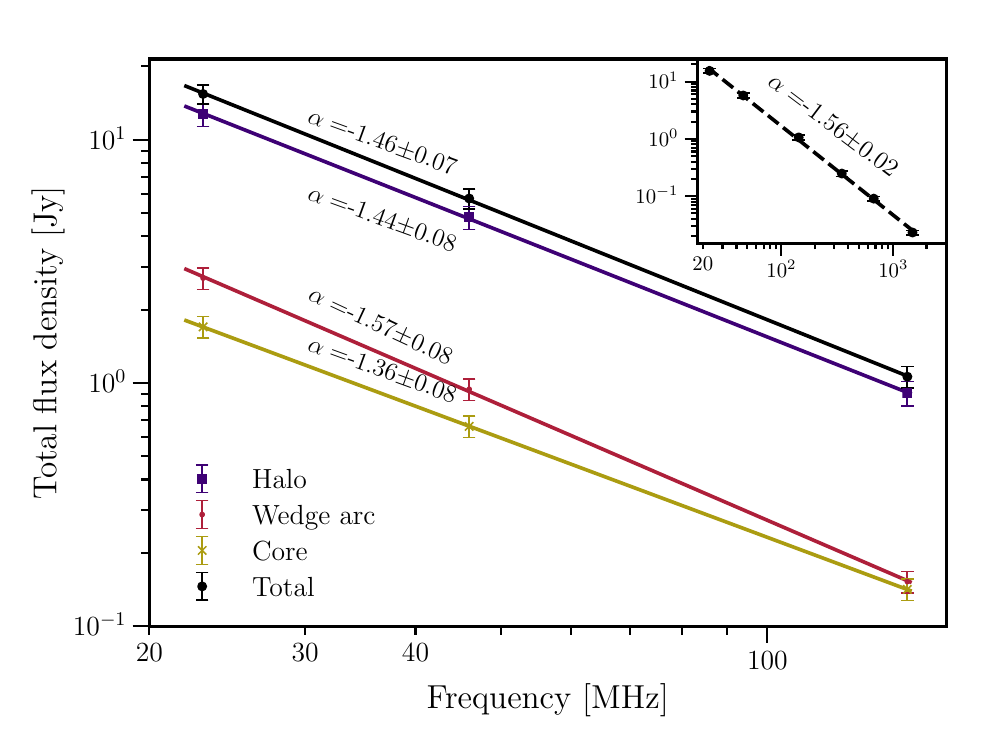}
    \caption{Integrated spectrum of the radio halo between $23$ and $144$ MHz. The three regions used to compute the spectrum are shown in Figure\,\ref{fig:radiobig}. The inset in the top-right corner shows the integrated spectrum with the addition of higher-frequency data from the recent study by \citet{Rajpurohit2023} and has the same axis labels as the main plot.}
    \label{fig:intspixhalo}
\end{figure}

We calculated the integrated spectral index of the radio halo between $23$ and $144$ MHz using the same regions from the recent higher frequency study from \citet{Rajpurohit2023} for both the subtraction of compact sources and the integration of the halo flux density, as well as the definition of the sub-regions `core' and `wedge' (see Fig\,\ref{fig:radiobig}). The resultant integrated spectrum is shown in Figure\,\ref{fig:intspixhalo}, where the inset shows our data together with the higher frequency measurements, where for a correct comparison accounting for the different baseline coverages, we filtered out baselines below 100$\lambda$ when measuring the halo flux. The resultant halo flux measurements are given in Table \ref{tab:halorelicflux}. The integrated spectral index follows a power-law of $-1.56 \pm 0.02$ over almost two orders of magnitude in frequency, from $24$ to $1500$ MHz. However, the spectral index of the core region is flatter than the overall radio halo, with $\alpha=-1.36\pm0.08$, while the spectral index of the `wedge' is slightly steeper at low frequencies. 

The integrated spectral index agrees within two standard deviations with the spectrum measured at higher frequencies ($\alpha=-1.63\pm0.03$; \citealt{Rajpurohit2023}), with no evidence for spectral curvature. This differs from the curved spectra observed in the radio halos of other clusters with wide frequency coverage, such as the Coma cluster, MACS J0717.5+3745 or Abell S1063 \citep{Thierbach2003,Rajpurohit2021,Xie2020}. 
We note that for our measurements we have ensured a consistent \textit{uv}-min in wavelength units for all datasets when imaging. This is highly important, as for example the halo is significantly brighter, by a factor of $\sim2$, at LOFAR frequencies in images without inner \textit{uv} cuts, indicating the presence of large-scale emission that is only detected with the shortest LOFAR baselines. Whilst our measurements integrated the pixels within a given region, we note that if we repeat the exercise but using the flux from the best-fit spherical models of the radio halo surface brightness (see Section\,\ref{CH4:gammaray} and Appendix \ref{CH4:halofit}) we find consistent results for the spectral index measurements. 

\begin{figure}[thb]
    \centering
    \includegraphics[width=1.0\columnwidth]{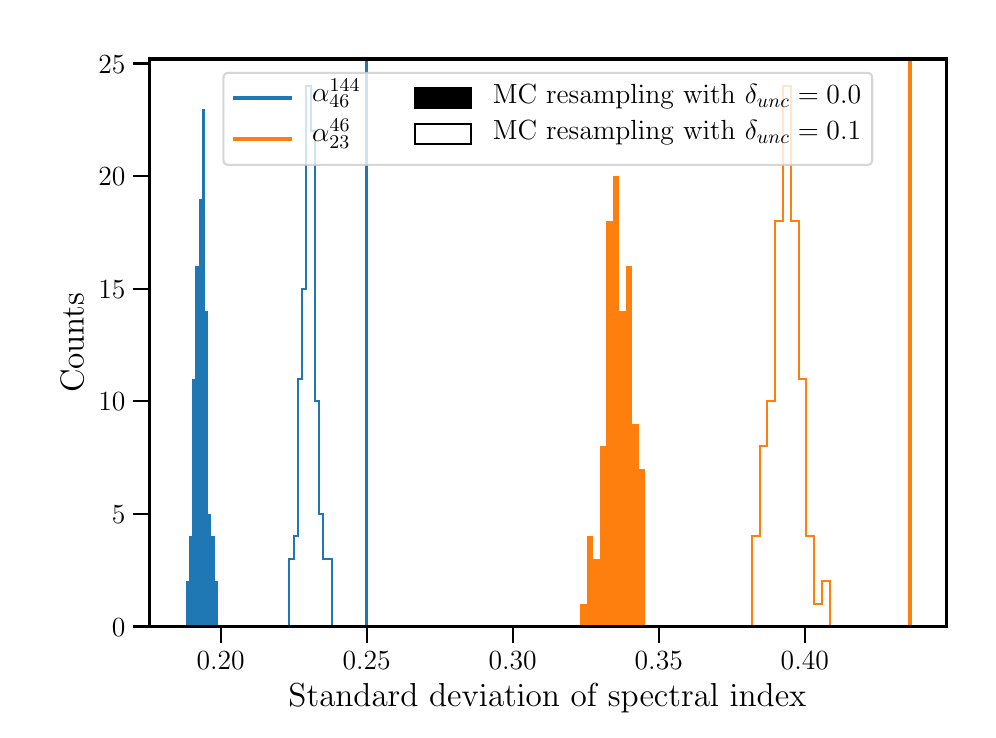}
    \caption{Observed standard deviation (vertical line) of the spectral index in the halo region, in comparison to the standard deviation expected purely from thermal noise ($\delta_\mathrm{unc}=0.0$) and an additional noise source that is 10\% of the surface brightness ($\delta_\mathrm{unc}=0.1$).}
    \label{fig:spixfluct}
\end{figure}

{To investigate the spatial distribution of the spectral index, we calculated the standard deviation from the halo region of the spectral index maps of the 23-46 and 46-144 MHz bands. These are std($\alpha_{23}^{46})=0.43$ and std($\alpha_{46}^{144})=0.25$, respectively, while the median uncertainties in the spectral index map in the radio halo region were found to be $0.36$ and $0.21$ respectively. However, the spectral index uncertainty varies slightly across the radio halo, so to check whether the observed values are higher than the fluctuations expected from noise, we performed the following Monte Carlo analysis. First, we calculated versions of the spectral index uncertainty maps with the only source of noise being background RMS fluctuations, as a systematic flux scale uncertainty would not contribute to spatial variations. Then, we assumed a constant spectral index map across the halo, and re-sampled every pixel in the constant map from a Gaussian with a standard deviation equal to the uncertainty in that location of the uncertainty map. This process was repeated 100 times, and the standard deviation of the re-sampled maps is then a good estimate of the level of fluctuations caused by noise. Figure \ref{fig:spixfluct} shows the comparison between the observed spectral index fluctuations and the magnitude of the fluctuations expected from noise. The observed standard deviation of the spectral index cannot be accounted for by map noise only.}

However, calibration or deconvolution errors might contribute noise as well. Therefore, we repeated this process assuming an additional noise source that is a fraction of the surface brightness, until we reach the observed standard deviation in the spectral index. We find that at least a 13\% surface brightness uncertainty would be needed to explain the observed standard deviation of the spectral index. Such large errors seem unlikely, given the fact that the radio halo in Abell 2256 is detected at high significance in all maps (see Fig. \ref{fig:radiolowres}), and LOFAR HBA observations have been shown to reliably recover more than 90\% of the flux of radio halos, even up to the scale of the radio halo in Abell 2256 \citep[e.g.][]{Shimwell2022,Bruno2023}. To estimate the level of noise added by calibration or deconvolution issues in the LBA maps, we calculated the flux density of the radio halo per self-calibration round and found that the results were very stable. The radio halo flux density varied on the order of 1\% between self-calibration rounds. It is thus unlikely that there are deconvolution errors that are causing additional surface brightness fluctuations larger than 10\%. We conclude that the spectral index shows excess scatter over the noise, likely caused by physical fluctuations in the emitting region.

\begin{figure}[thb]
    \centering
    \includegraphics[width=1.0\columnwidth]{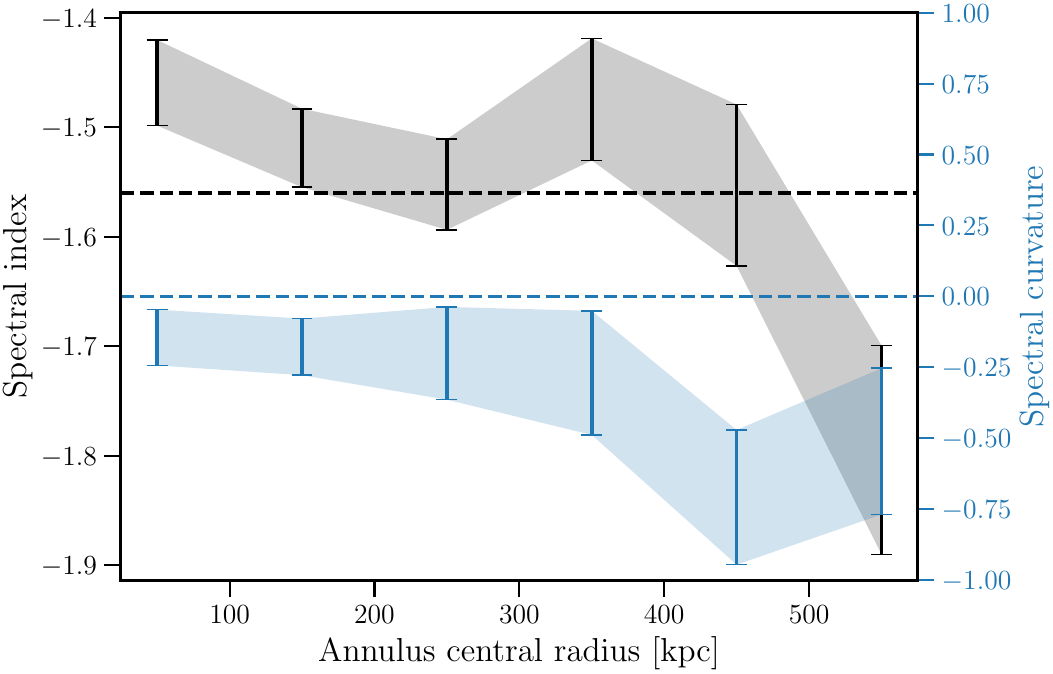}
    \caption{Spectral index (black) and curvature (blue) profile as a function of radius from the halo centre, computed in concentric annuli from a least-square fit to the $23$, $46$ and $144$ MHz data. The value of the integrated spectral index and curvature of the radio halo is indicated by the dashed lines.}
    \label{fig:spixradius}
\end{figure}

Higher frequency data on Abell 2256 has shown that the spectral index also has a radial trend, with steeper spectra towards the outskirts \citep{Rajpurohit2023}. The low-frequency spectral index maps in Figure\,\ref{fig:radiofullspix} do not show a clear radial trend when analysed in a comparable way. To quantify this, we calculated the spectral index by fitting the 23, 46 and 144 MHz flux densities in radial bins using concentric annuli. The best-fit spectral index, with the one-sigma uncertainty, is shown as a function of radius in Figure\,\ref{fig:spixradius}. Although there is a hint of steepening in the last radial bin and flattening in the core, the data are consistent with a constant spectral index as a function of radius, given the large error bars. 

\begin{figure}[thb]
    \centering
    \includegraphics[width=1.0\columnwidth]{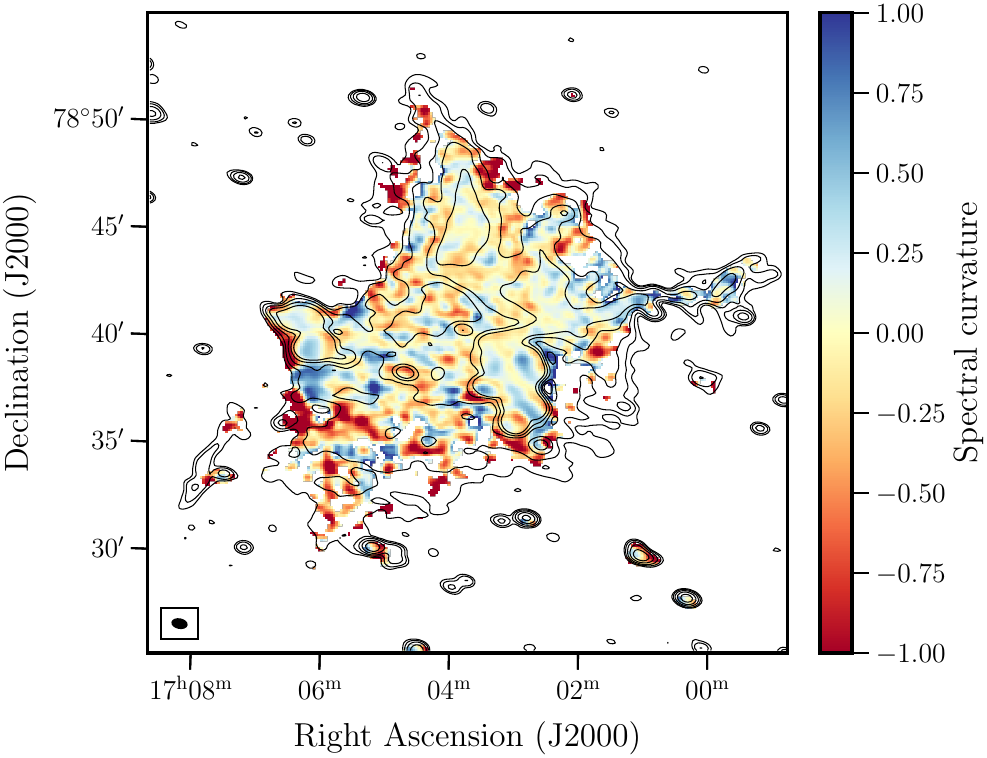}
    \caption{Spectral index curvature map of Abell 2256, calculated as $\alpha_{23\mathrm{MHz}}^{46\mathrm{MHz}}-\alpha_{46\mathrm{MHz}}^{144\mathrm{MHz}}$, where both spectral index maps are made with a common resolution of 39$^{\prime\prime}\times24^{\prime\prime}$. Blue regions indicate spectral steepening towards higher frequencies and red regions indicate flattening at higher frequencies. The background contours show the 144 MHz intensity smoothed to the same resolution.}
    \label{fig:curv}
\end{figure}

A superposition of curved spectra could result in a single power-law spectrum when integrated over the entire radio halo, as observed for example in Abell 2744 \citep{Rajpurohit2021b}. 
To investigate possible spatial variation of the curvature, we computed the spectral curvature map from the spectral index maps as follows
\begin{equation}
\alpha_{23\mathrm{MHz}}^{46\mathrm{MHz}}-\alpha_{46\mathrm{MHz}}^{144\mathrm{MHz}}.
\end{equation}
This spatial distribution of spectral curvature is shown in Figure\,\ref{fig:curv}, and the corresponding uncertainty map in Figure\,\ref{fig:curv_err}. The median uncertainty across the radio halo is 0.39 while the standard deviation of the measured curvature across the radio halo is 0.52. Using the same Monte Carlo test as above, we find that setting a level of 10\% noise fluctuations due to calibration or deconvolution errors produces a lower standard deviation in curvature (0.430 +/- 0.005) than is observed. Thus, the spectrum locally exhibits a convex or concave shape in different regions of the radio halo, likely due to physical fluctuations in the emitting region. Due to large uncertainties, we cannot say if the spectral curvature variations show a radial dependence, as Figure\,\ref{fig:spixradius}, is consistent with a straight line.

In summary, the radio analysis of the halo in Abell 2256 reveals that the integrated spectrum of the halo is consistent with a steep power-law. However, we also find evidence of spectral index and curvature variations that do not follow a radial profile, indicating a complex and inhomogeneous environment.

\subsection{Radio shock}\label{CH4:radioshock}
Abell 2256 hosts one of the clearest examples of filamentary radio emission inside a radio shock \citep[][]{Clarke2006,Brentjens2008,Owen2014}. Usually, such filaments are observed at gigahertz frequencies only due to the high resolution required but here we present the first case where filamentary radio emission is observed in a radio shock down to at least 23\,MHz. In the total intensity images shown in Figures \ref{fig:radiobig} and \ref{fig:radiofullres} the well-known filaments inside the radio shock (source G and H) are still clearly seen at 144 and 46 MHz, while the 23 MHz image shows only the brightest larger filaments. 

\begin{figure}[thb]
    \centering
    \includegraphics[width=1.0\columnwidth]{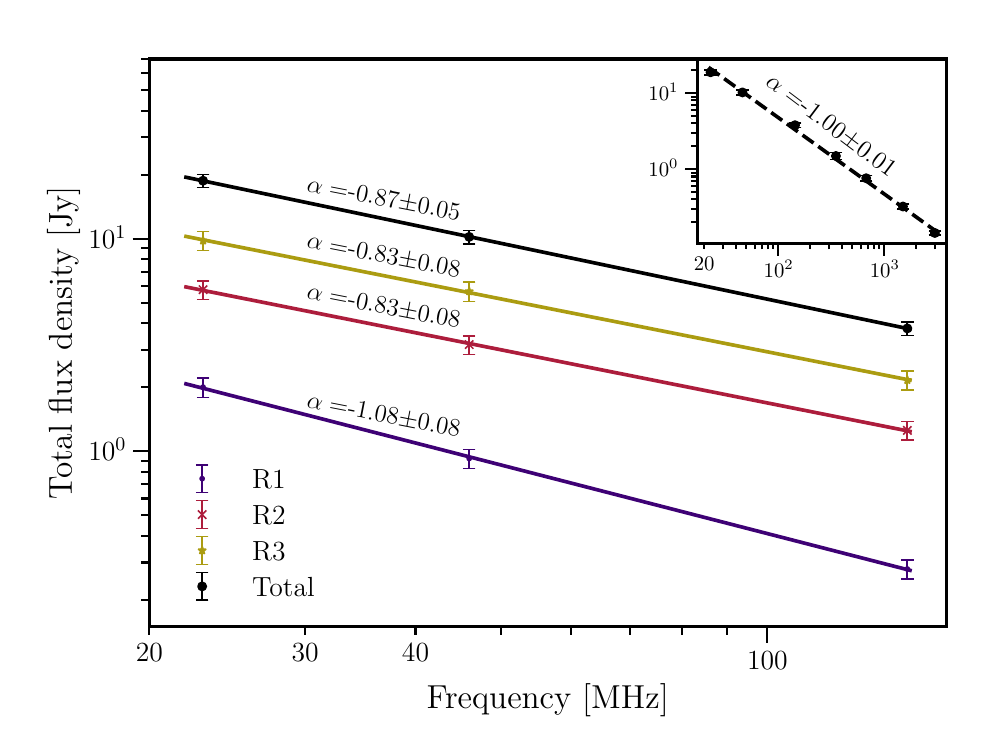}
    \caption{Integrated spectrum of the radio shock between $23$ and $144$ MHz. The three regions used to compute the spectrum are shown in Figure\,\ref{fig:spixrelic}. The inset in the top-right corner shows the integrated spectrum with the addition of higher-frequency data from the recent study by \citet{Rajpurohit2022b} and has the same axis labels as the main plot.}
    \label{fig:intspixrelic}
\end{figure}

\begin{figure}[thb]
    \centering
    \includegraphics[width=1.0\columnwidth]{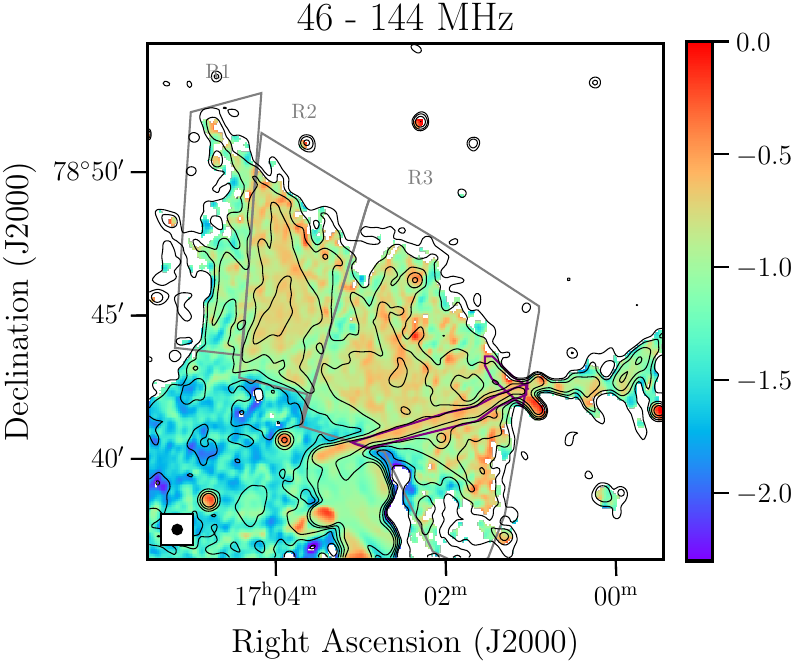}
    \caption{Spectral index map of the radio shock between $46$ and $144$ MHz made at a common resolution of 20$^{\prime\prime}\times20^{\prime\prime}$. The three regions used to compute the integrated spectral index of the shock are shown in grey, where the contribution from the bright narrow-angle-tailed sources is subtracted, shown in purple. The contours indicate the LOFAR 144 MHz total intensity at [3,6,12,24,48] times the background noise.}
    \label{fig:spixrelic}
\end{figure}

The integrated spectrum of the radio shock between 23 and 144 MHz is plotted in Figure\,\ref{fig:intspixrelic}, where we have divided the shock into three sub-regions (shown in Figure\,\ref{fig:spixrelic}) to allow us to study the spectral steepening from west to east across the radio shock that was noticed at higher frequencies by \citet{Clarke2006} and \citet{Rajpurohit2022b}. The radio shock flux measurements are given in Table \ref{tab:halorelicflux}. In agreement with these previous studies, Figure\,\ref{fig:intspixrelic} shows that the westernmost region R1 has a steeper spectrum than regions R2 and R3, with R1 showing $\alpha=-1.08\pm0.07$ and R2 and R3 showing $\alpha=-0.84\pm0.07$ and $\alpha=-0.83\pm0.08$, respectively. We note that the uncertainties on the spectral index measurements are dominated by the systematic uncertainty in the flux density scale, and thus have a strong spatial correlation while the statistical uncertainty is on the order of $0.02$. The total integrated spectrum, when combined with higher frequency data from the recent study by \citet{Rajpurohit2022b}, agrees with a straight power-law with $\alpha=-1.00\pm0.01$ from 24 to 3000 MHz. 
However, at low frequencies, the contribution of the radio halo flux to the region of the radio shock might become significant and cannot be easily separated in the images. We can estimate the contribution using the spherical halo models that were fit in Appendix \ref{CH4:halofit}. Assuming the radio halo is spherically symmetric, we find that the halo contributes 10\%, 26\% and 39\% of the total radio shock flux in the 144, 46 and 23 MHz images respectively. However, subtracting this contribution only flattens the spectrum marginally. With the subtraction of the estimated radio halo flux from the radio shock region, we find that the radio shock spectrum between 24 and 3000 MHz still follows a power-law with $\alpha=-0.95\pm0.01$. 

The spectral index trend across the radio shock can also be seen in the spectral index map at 20$^{\prime\prime}$ between 46 and 144 MHz, shown in Figure\,\ref{fig:spixrelic}. This spectral index map indicates steepening from the southwest towards the northeast side of the radio shock, where we see preferentially emission with $\alpha<-1$. In contrast, the west side of the radio shock shows flatter spectrum emission, with $\alpha>-1$. 

\subsection{AGN related emission}
Abell 2256 also hosts a large number of complex radio sources, that appear to be either directly related to AGN or associated with (revived) fossil AGN plasma \citep{Weeren2009}. Fossil plasma sources typically show very steep spectra that are often curved at high (GHz) frequencies \citep[e.g.][]{Mandal2020}. In the case of Abell 2256, there are various (candidate) fossil plasma sources. 

First, the sources labelled AG+AH and AI were discovered in \citet{Weeren2009}, where they showed spectral indices at frequencies higher than 140 MHz of $\alpha<-1.95$ and $\alpha<-1.45$ respectively. The possibility was raised that both sources are revived fossil plasma sources, although AG+AH might also simply be old AGN emission from the long, tailed radio galaxy. This scenario is supported by the high-resolution radio images of Figures \ref{fig:radiobig} and \ref{fig:radiofullres}, where AG+AH seems to be connected to the long Mpc-sized tailed radio source C. In the high-resolution HBA image, we also clearly observe for the first time `ribs' coming off the source AG+AH. These are reminiscent of the ribs seen in the radio tail dubbed T3266 in Abell 3266 as observed with the MeerKAT telescope \citep{Knowles22,Riseley2022,Rudnick2021}

\begin{figure}[thb]
    \centering
    \includegraphics[width=1.0\columnwidth]{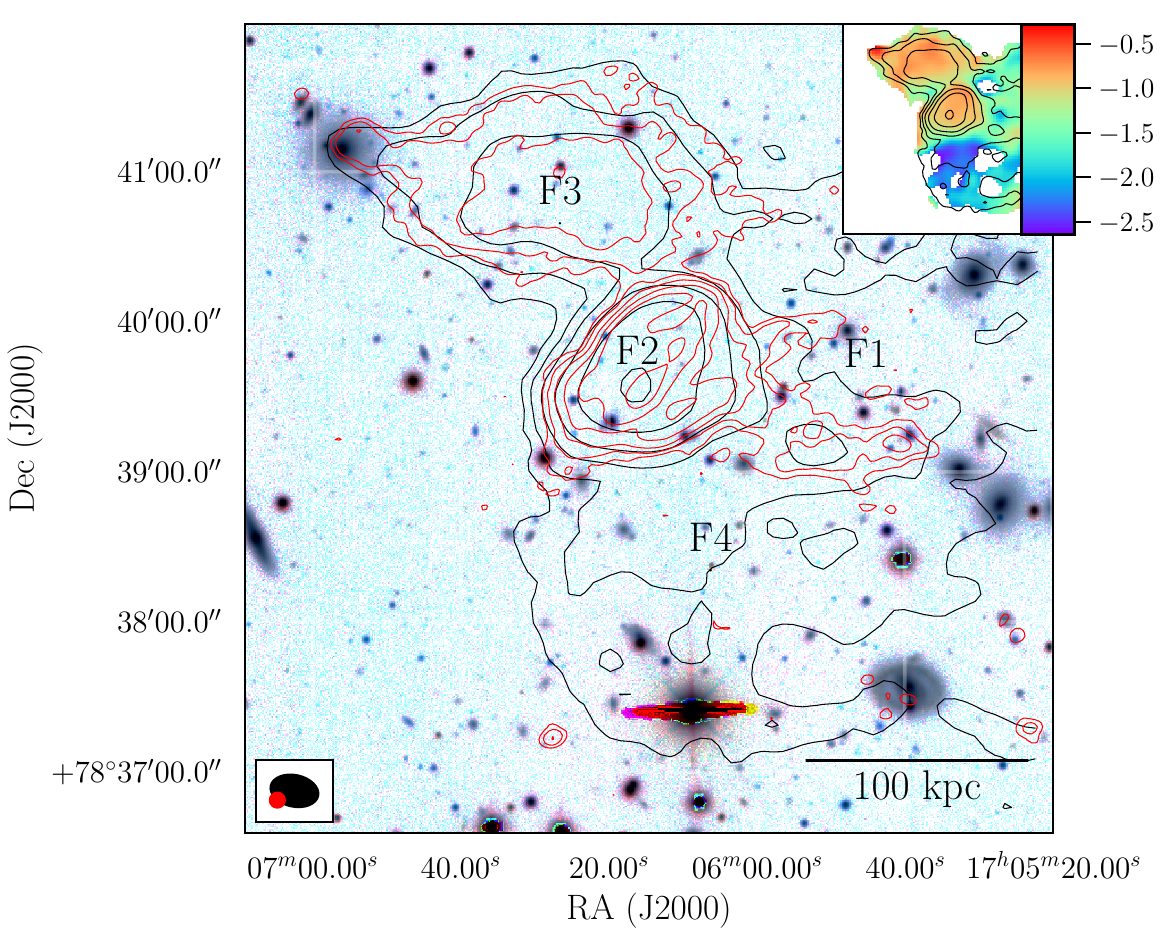}
    \caption{The F-complex in Abell 2256, shown as the overlay of LOFAR 144 MHz $6^{\prime\prime}$ (red) and LOFAR 46 MHz $19^{\prime\prime}$ (black) contours on the \textit{grz} optical filters from the Legacy Survey \citep[][]{Dey2019}, with inverted colours for visibility. The restoring beams are indicated in the lower left corner, and contours are drawn at [5,10,20,40] times the background noise level $\sigma$. The inset plot shows the spectral index map between 46 and 144 MHz at the common $19^{\prime\prime}$ resolution with the same LOFAR 46 MHz contours, where F4 is visible as the blue steep region of emission.}
    \label{fig:opticalFcomplex}
\end{figure}

Second, there is the F complex of sources, also discussed by \citet{Bridle1976,Bridle1979,Rottgering1994,Brentjens2008} and \citet{Owen2014}. The F complex of sources is located on the west side of the radio halo, and consists of three components, F1, F2 and F3. The narrow-angle tailed source F3 is clearly associated with a cluster member (\citet{Fabricant1989} galaxy 122), situated at the eastern tip of the radio source \citep{Owen2014}, as shown in Figure\,\ref{fig:opticalFcomplex}. However, the nature and origin of the other two sources are still unclear. One possibility is that F1 and F2 are also related to the same galaxy as F3, but another possibility is that F1 and F2 consist of fossil radio plasma from previous episodes of AGN activity (possibly from F3) that is compressed somehow by interactions in the ICM \citep{Weeren2012Abell}. 

The 23 MHz data shows that F2 and F3 are more extended than previously reported at higher frequencies \citep{Brentjens2008,Owen2014}. The radio emission of F3 seems to fade into the wedge arc of the radio halo, indicating a possible connection between the tailed radio source and the halo arc. 
Interestingly, we observe no clear spectral index gradient across F1-F3. 
Additionally, we detect a new, very steep, region just below the F complex, co-spatial with the radio halo. It is clearly seen as a bright region in the 46 MHz contours shown in Figure\,\ref{fig:opticalFcomplex} and shows a spectral index of $\alpha<-2$ in the 46-144 MHz spectral index map (Fig. \ref{fig:radiofullspix}). This seems like a fossil plasma source due to the extreme steepness of the spectrum, and could possibly be associated with the F complex as well. We, therefore, label it F4 in this study. The optical overlay, Figure\,\ref{fig:opticalFcomplex}, shows that the 46 MHz contours seem to originate from the cluster galaxy MCG+13-12-020 at 17h05m39.5s +78d37m34.2s the south-west, which agrees with the spectrum flattening spatially towards this galaxy, implying a possible optical host.  

\begin{figure}[thb]
    \centering
    \includegraphics[width=1.0\columnwidth]{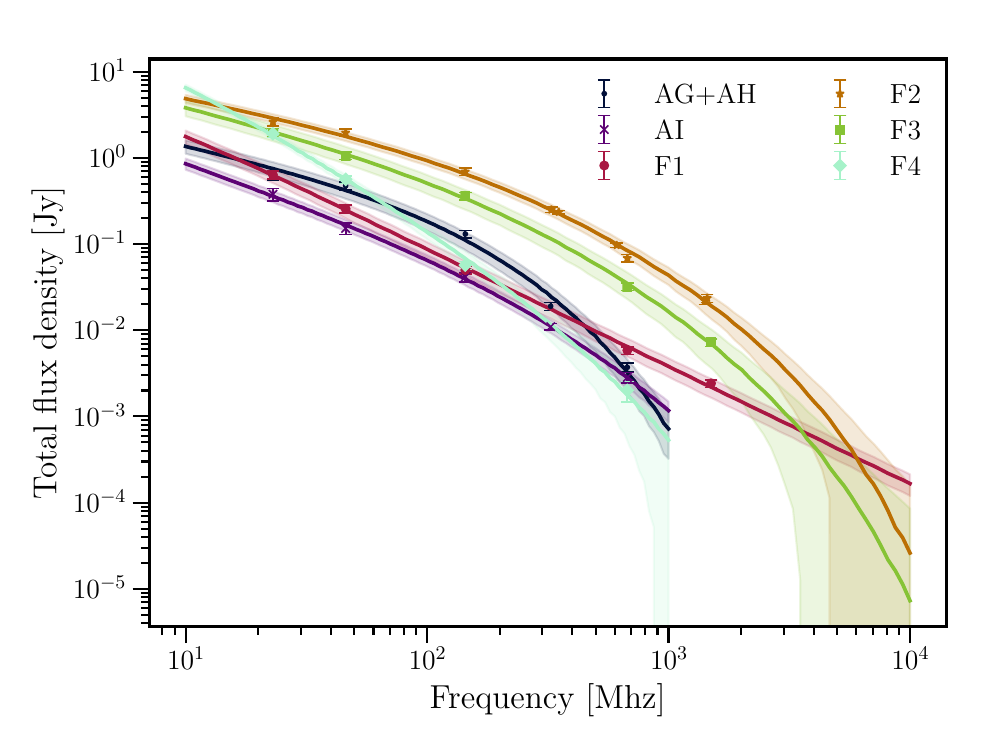}
    \caption{Spectra of fossil plasma sources down to 23 MHz adding literature data above 150 MHz where available. The flux density measurements and best-fit parameters of the synchrotron ageing models are given in Table \ref{tab:AGNsources}.}
    \label{fig:AGNspectra}
\end{figure}

As lower energy electrons cool less efficiently through synchrotron and inverse Compton radiation, our low-frequency data allows us to probe the aging of the observed emission. By fitting their spectra (see Fig\,\ref{fig:AGNspectra}) with simple synchrotron ageing models, we estimated the ages of AG, AH, AI and F1-F4, adding high-frequency data from the literature where possible. All sources except F1 and F4 show spectral flattening towards lower frequencies, indicating that we do not observe the break frequencies of F1 and F4, which are likely below $23$ MHz. 

We used \textit{synchrofit}\footnote{https://github.com/synchrofit} \citep{Quici2022} to fit standard synchrotron models to the various AGN-related sources in Abell 2256. We fit a continuous injection \citep[KGJP;][]{KomissarovGubanov1994} model to the curved spectrum sources. The model has three free parameters: the injection index $s=1-2\alpha_\mathrm{inj}$, where $\alpha_\mathrm{inj}$ is the radio spectral index upon injection, the break frequency after which the spectrum steepens, and the remnant fraction (i.e. the fraction of time the source is `off'). Following the minimum energy condition as calculated in \citet{Brentjens2008} which follows the \citet{Beck2005} formula, we assume a tangled magnetic field with a strength of 7 $\mu$G for 
the F complex. Doing the calculation for AG+AH and AI gives lower values of the minimum energy magnetic field strength around $\sim$3 $\mu$G, but we assume $7\mu$G as well to give a conservative age estimate. We note that the maximum age estimate is obtained for $B=B_\mathrm{CMB}/\sqrt{3}$ \citep[e.g.][]{Stroe2014}, which results in $1.8\mu$G at the redshift of Abell 2256. The resulting spectral ages, best-fit injection indices and break frequencies for the AGN-related sources are given in Table \ref{tab:AGNsources}. 

The straight spectrum of F1 over multiple decades in frequency indicates the source is likely still being energised and we are observing the spectrum above the break frequency. For a simple continuous injection model, the spectrum would consist of two power-laws with  $\alpha=\alpha_\mathrm{inj}-0.5$ after the break frequency \citep{Pacholczyk1970}. The best-fit spectral index of F1 was found to be $\alpha=-1.36\pm0.03$, implying a radio injection index of $\alpha_\mathrm{inj}=-0.86\pm0.03$. 
For source F4 the simple continuous injection model does not fully work, because it would imply an injection index of $\alpha_\mathrm{inj}=0.5-1.9\pm0.1=-1.4\pm0.1$, which is much steeper than typical injection indices ($>-1$). Thus we are likely observing the exponential steepening of the spectrum of F4, implying relativistic particles are not continuously injected.

\begin{table*}[htb]
\centering
\caption{Flux density measurements and best-fit synchrotron model parameters of the AGN-related sources in Abell 2256. }
\label{tab:AGNsources}
\resizebox{\textwidth}{!}{%
\begin{tabular}{@{}lllllllll@{}}
\toprule
Source &
  $S_\mathrm{144MHz}${[Jy]} &
  $S_\mathrm{46MHz}${[Jy]} &
  $S_\mathrm{23MHz}${[Jy]} &
  Model &
  $\alpha_\mathrm{inj}$\tablefootmark{a} &
  Break frequency\tablefootmark{a}{[MHz]} &
  Remnant Fraction\tablefootmark{a} &
  Age\tablefootmark{b}{[Myr]} \\ \midrule
AG+AH & 0.13 $\pm$ 0.01 & 0.47 $\pm$ 0.05 & 0.64$\pm$0.08 & {KGJP} & -0.55$\pm$0.02 & 113$\pm$12 & 0.57$\pm$0.05 & 197$\pm10$ \\
AI    & 0.04 $\pm$ 0.01 & 0.15 $\pm$ 0.02 & 0.38$\pm$0.06 & KGJP & -0.91$\pm$0.02 & 242$\pm$82 & 0.33$\pm$0.04 & 135$\pm30$ \\
F1    & 0.05 $\pm$ 0.01 & 0.26 $\pm$ 0.03 & 0.64$\pm$0.07 & CI\tablefootmark{c} & -0.86$\pm0.02$ & $<23$ & - & $>437$ \\
F2    & 0.69 $\pm$ 0.07 & 1.96 $\pm$ 0.20 & 2.62$\pm$0.26 & KGJP & -0.53$\pm$0.01 & 129$\pm$22 & 0.24$\pm$0.05 & 185$\pm15$ \\
F3    & 0.36 $\pm$ 0.04 & 1.06 $\pm$ 0.11 & 1.97$\pm$0.20 & KGJP & -0.66$\pm$0.02 & 145$\pm$24 & 0.25$\pm$0.05 & 174$\pm14$ \\
F4    & 0.06 $\pm$ 0.01 & 0.56 $\pm$ 0.06 & 1.91$\pm$0.19 & CI\tablefootmark{c} & {-1.4$\pm$0.1} & $<23$ & - & $>437$ \\ \bottomrule
\end{tabular}%
}
\tablefoot{\tablefoottext{a}{The error bars only reflect the statistical uncertainties from the fit.}\tablefoottext{b}{For the age estimate, a conservative magnetic field value of $7 \mu$G was assumed, following \citet{Brentjens2008}. For a lower magnetic field strength, the age would increase.}\tablefoottext{c}{F1 and F4 are consistent with a simple power-law spectrum without a break frequency observed.}}
\end{table*}

\section{Radio - Gamma-ray comparison}\label{CH4:Gammaray}
Nearby clusters such as Abell 2256, whose radio halo exhibits an ultra-steep spectrum, are expected to generate gamma-ray flux in the Fermi-LAT energy band if the halo is generated by secondary particles from hadronic interactions \citep{Brunetti2009}. They are therefore ideal candidates to constrain the contribution of secondary electrons from hadronic interactions to the cosmic ray electron population. In this section, we combine our LOFAR data with upper limits from Fermi-LAT data to test a purely hadronic origin of the halo. 

\subsection{Theoretical framework}\label{CH4:theory}
{To study the contribution of the hadronic interactions to the radio halo in Abell 2256, in this section, we model hadronic interactions of cosmic ray protons with thermal ions. Assuming spherical symmetry, we obtained the thermal properties of Abell 2256 using X-ray data from ACCEPT, ROSAT \citep{Truemper1993,Eckert2012,Eckert2013b,Eckert2013a}, and the Sunyaev-Zel’dovich data from Planck \citep{PSZ2}. We fit a gNFW profile \citep{Nagai2007} for the pressure and a simple $\beta$-model for the gas density as a function of radius ($R$). The best-fit parameters for the beta model were found to be $n_\mathrm{th}(0) = 3\times10^{-3}$ cm$^{-3}$, $r_c$ = 341 kpc, and $\beta=0.77$
in the standard $\beta$-model given by 
\begin{equation}
n_\mathrm{th}(R) = n_\mathrm{th}(0) \left[ 1 + \left( \frac{R}{r_c} \right)^2 \right]^{-3\beta/2}
\end{equation}
We verified that the fits also closely match the X-ray data from the Archive of Chandra Cluster Entropy Profile Tables \citep[ACCEPT;][]{Donahue2006,Cavagnolo2009} out to $R_\mathrm{500}$. 
}

{For the non-thermal properties, we assumed a power-law distribution of cosmic-ray proton density that follows the thermal plasma distribution $n_{th}$ as follows:}
\begin{equation}
N_{CRp}(p_p,R)  = C_p [ n_{th}(R) kT(R) ]^a p_p^{-s},
\label{eq:CRspectrum}
\end{equation}
where $C_p$ is a constant, $n_{th}$ and $kT$ denote the thermal gas density and temperature as functions of radius $R$, and $p_p$ denotes the momentum of the protons, following a power-law distribution with index $s$, down to a conservative momentum cut-off $p_{p,\mathrm{min}} \sim 0.1 mc$ (approximately 30 times larger than that of the thermal protons). { A power-law momentum distribution for cosmic-ray protons is routinely assumed in the literature as a result of acceleration mechanisms, and agrees with the power-law radio spectrum found for the halo down to low frequencies}. The proportionality between the cosmic ray proton energy density and the thermal plasma energy density ($W_\mathrm{CRp}\propto {W_\mathrm{th}}^{a}$) is parameterised by $a$, which is constrained from the Abell 2256 radio halo intensity profile in the next section (Fig. \ref{fig:haloprofile}). 

The collisions between CRp\footnote{For CRp with momentum above the threshold $p> p_{thr}$; the threshold kinetic energy being 289 MeV \citep[e.g.][]{Norbury2009}} and thermal protons create pions (denoted by $\pi$) that decay into $\gamma$-rays, electrons/positrons and neutrinos \citep[e.g.][]{BlasiColafrancesco1999}. The spectrum of secondary electrons is calculated as in \citet{Brunetti2017}, assuming stationary conditions. First, the injection spectrum of electrons and positrons is given {by numerical integration:}
\begin{equation}
 \begin{aligned}
Q_{e}^{\pm}(p_e)=& \frac{8 \beta_{\mu}^{\prime} m_{\pi}^{2} n_\text{th} c^{2}}{m_{\pi}^{2}-m_{\mu}^{2}} \int_{E_{\min }} \frac{d E_{\pi}}{E_{\pi}}
\int_{p_{*}} \frac{\text{d} p_p}{\bar{\beta}_{\mu}} \beta_{p} N_{CRp}(p_p) \\
& \times \frac{\text{d} \sigma^{\pm}}{\text{d} E}\left(E_{\pi}, E\right) F_{e}\left(E_{e}, E_{\pi}\right),
\end{aligned}
\label{eq:electronspectrum}
\end{equation}
{where $p_e$ is the electron momentum,} $E$ is the proton energy (otherwise $E_i$ is the energy of species $i$), $m_\pi$ and $m_\mu$ are the pion and muon masses, $\beta_{\mu}^{\prime}=0.2714$, $E_\mathrm{min}=2E_e m_\pi^2/(m_\pi^2+m_\mu^2)$, $p_* = max\{p_{th}, p_{\pi}\}$, $\bar{\beta}_{\mu}=\sqrt{1-m_\mu^2/\bar{E_\mu^2}}$, $\bar{E_\mu}=1/2E_\pi(m^2_\pi-m^2_\mu)/(\beta_{\mu}^{\prime}m_\pi^2)$ 
and $d\sigma^{\pm}/dE$ is the differential inclusive cross-section for the production of neutral and charged pions. This cross-section is calculated by combining four energy ranges, following  \citet{Brunetti2017} and references therein. Finally, $F_{e}\left(E_{e}, E_{\pi}\right)$ is given in \citet[][]{BrunettiBlasi2005} below Eq. 36. The resulting steady-state distribution of the secondary electrons is then calculated as 
\begin{equation}\label{eq:Ne}
 N_e^{\pm}(p_e)=\frac{1}{\left|\frac{\text{d}p_e}{\text{d}t}\right|_\mathrm{rad} + \left|\frac{\text{d}p_e}{\text{d}t}\right|_\mathrm{C}}\int_p Q_e^{\pm}(x)dx,
\end{equation}
where $\left|\text{d}p_e/\text{d}t\right|_\text{i}$ denotes radiative ($i$=rad) and Coulomb ($i$=C) losses, from \citet{Brunetti2017}. {The steady-state distribution is a good assumption for galaxy clusters, as CRp are confined to the cluster for many Gyr \citep[e.g.][]{BrunettiJones2014}, and the timescale of p-p collisions is much larger than the relatively short lifetime of synchrotron-emitting (GeV) electrons ($\sim 10^8$ yr, e.g. \citealt{Weeren2019}). Thus within a few cooling times, the electron spectrum will reach a steady state balance between injection and cooling.}

\noindent
{These electrons will generate a synchrotron emissivity obtained from the following numerical integration:}
\begin{equation}
j_{syn}(\nu) = \sqrt{3}
\frac{e^3}{m_e c^2} \int_0^{\pi/2} d\theta \sin^2\theta
\int N_e^{\pm}(p_e) F\left(\frac{\nu}{\nu_c}\right) dp_e,
\label{syn-emissivity}
\end{equation}
{where $e$ is the elementary charge, $m_e$ the electron mass, $c$ the speed of light, $\nu_c$ is the synchrotron critical frequency and $F$ is the synchrotron Kernel \citep[e.g.][]{RybickiLightman1979}. The pitch angle $\theta$ between the magnetic field and the electron velocity is assumed to be randomly distributed.}

{Assuming a power-law distribution of CRp as in Eq. \ref{eq:CRspectrum}, the spectrum of the synchrotron emissivity can be approximated with a power-law in the form $j_{syn}(\nu) \propto \nu^{\alpha}$, with $\alpha \simeq (1-s)/2$}.\footnote{ We note that Coulomb losses may generate a flattening in the spectrum of CRp at lower energies, which may induce a corresponding flattening in the synchrotron spectrum generated by secondary electrons at low frequencies. In the case of galaxy clusters, this effect is expected to be significant only in the cores, where densities and magnetic field strengths are higher. However, the low-frequency data excludes the possibility of flattening down to energies of $E \sim 3-10$ GeV and only a small fraction of the gamma rays are produced in the core, thus for simplicity we neglect this effect.}
The gamma-ray intensity from the decay of pions was then computed following \citet[][and ref. therein]{Brunetti2017}, with the injection rate of pions given by
\begin{equation}
Q_{\pi}^{\pm,0}(E_{\pi})= n_\mathrm{th} c 
\int_{p_{*}} dp_p N_{CRp}(p_p) \beta_p 
\frac{ d \sigma^{\pm, 0} }{ d E } (E_{\pi},E),
\label{q_pi}
\end{equation}
where $\pm$ and $0$ refer to charged and neutral pions, respectively, and $d \sigma^{\pm, 0}/d E$ is the differential inclusive cross section for their production, which is calculated in four different energy ranges as in \citet{Brunetti2017}. {The decay of neutral pions then generates an emissivity in the gamma-rays at the energy E$_\gamma$ in the form}
\begin{equation}
j_{\gamma}(E_{\gamma})= 2 E_{\gamma}\int_{E_{min}}^{E_{CRp,max}}
\frac{Q_{\pi}^0(E_{\pi})}{\sqrt{E_{\pi}^2 - m_{\pi}^2 c^4}}
dE_{\pi},
\label{eq:gammaraysspectrum}
\end{equation}
where $E_{min} = E_{\gamma} + m_\pi^2c^4/(4E_{\gamma})$.

{The synchrotron and gamma-ray emission can
be obtained through numerical integration of Eqs. \ref{eq:electronspectrum}--\ref{syn-emissivity} and \ref{q_pi}--\ref{eq:gammaraysspectrum}, respectively. As a useful reference (using a power-law approximation for the synchrotron emissivity), at a distance $r$ on the sky plane, the synchrotron and gamma-ray emission are proportional to:}
\begin{equation}\label{eq:Isynch}
{I_{\mathrm{syn}}(r)} 
\propto \int_\mathrm{LOS} \frac{RdR}{\sqrt{R^2-r^2}} n_\mathrm{th}^2(R) kT(R) \mathcal{F}(R) \frac{B^{1-\alpha}(R)}{B^2(R)+B^2_\mathrm{CMB}},
\end{equation}

{and} 

\begin{equation}\label{eq:gamma}
{I_{\mathrm{\gamma}}(r)} 
\propto \int_\mathrm{LOS} \frac{RdR}{\sqrt{R^2-r^2}} n_\mathrm{th}^2(R) kT(R) \mathcal{F}(R) ,
\end{equation}
{where we defined $\mathcal{F}(R)= \frac{W_\mathrm{CRp}(R)}{W_\mathrm{th}(R)}$ and $B$ refers to the {ICM magnetic field strength, with the CMB subscript referring to the cosmic microwave background equivalent magnetic field strength}.}
{The ratio of the synchrotron to gamma-ray luminosity is thus governed by the magnetic field profile as 
\begin{equation}
    \frac{L_\mathrm{syn}}{L_\mathrm{\gamma}} \propto \left< \frac{B(R)^{1-\alpha}}{B^2(R)+B^2_\mathrm{CMB}}\right>,
\end{equation}
where the brackets denote a volume average {(e.g. in the next section we integrate up to $R_\mathrm{500}$)} weighted for the distribution of CRp \citep{Brunetti2017}.}

{In our calculations,} the cluster magnetic field was assumed to follow the commonly used profile where the magnetic field energy density is proportional to the thermal gas energy density, as found for example for the Coma cluster \citep[][]{Bonafede2010} 
\begin{equation}\label{eq:Bfieldprofile}
B(R) = B_0 \left(\frac{n_\mathrm{th}(R)}{n_\mathrm{th}(0)}\right)^{0.5},
\end{equation}
where $n_\mathrm{th}(R)$ denotes the thermal electron density at radius $R$. The central magnetic field strength $B_0$ is not well-constrained for Abell 2256 \citep[e.g.][]{Ge2020}, so was left as a free parameter.

{In summary, with reasonable assumptions on the magnetic field profile and cosmic ray proton spectrum plus measurements of the cluster thermal density, temperature profile, and synchrotron luminosity, we can estimate the expected gamma-ray luminosity from hadronic interactions in Abell 2256. We additionally assumed spherical symmetry and homogeneous and stationary conditions for simplicity. 
The results may be influenced by non-homogeneous conditions within the intra-cluster medium. For instance, in a similar analysis conducted on the Coma cluster, \citet{Brunetti2012} demonstrated that for an additional turbulent component of the magnetic field (where 
$\left<{B} + \delta B\right>_\mathrm{Volume}= {B}$ and $\left<{B} + \delta B \right>^2_\mathrm{Volume}= {B}^2 + \delta B^2$), the radio/gamma-rays ratio changes by less than a factor of 2 compared to that in a homogeneous medium, even in the extreme scenario where $\delta B^2 \sim {B}^2$. Thus, the main conclusions are not expected to change significantly despite potential variations in intra-cluster medium conditions.
}

\subsection{Gamma-ray upper limits}\label{CH4:gammaray}

The LOFAR observations of the radio halo in Abell 2256 constrain the spatial distribution (i.e. $a$) and number density of cosmic ray protons of the purely hadronic model. We obtained the spatial distribution CRp from the brightness profile of the radio halo, which should follow Equation \ref{eq:Isynch}. We modelled the radio surface brightness using an MCMC halo-fitting code \citep{Boxelaar2021}. We masked the regions where the halo is seen in projection with either AGN or the large radio shock, as shown in Figure\,\ref{fig:halofit}. We assumed a simple spherically symmetrical model commonly used for radio halos where $I(r) = I_0 \exp(-r/r_e)$ \citep[e.g.][]{Osinga2021,Weeren2021,Edler2022}. The resulting fits are shown in Appendix \ref{CH4:halofit} with the best-fit model parameters given in Table \ref{tab:halofit}. We found similar values for the $e$-folding radius of $\sim 200$ kpc at the three different frequencies, which is consistent with the finding in Section \ref{CH4:radiohalo} that the spectral index is constant as a function of radius.

\begin{figure}[thb]
    \centering
    \includegraphics[width=1.0\columnwidth]{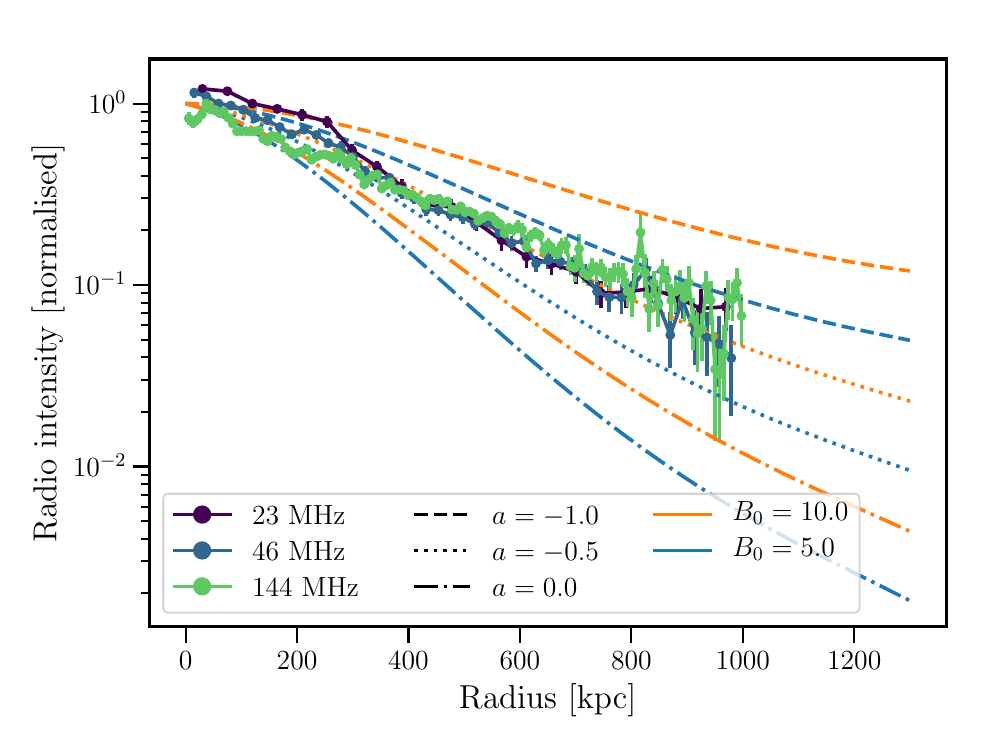}
    \caption{Observed (points) and modelled (lines) radio halo synchrotron intensity profiles in Abell 2256. The parameter $B_0$ denotes the central magnetic field strength, and $a$ the proportionality between cosmic ray energy density and thermal energy density. The details of the models are explained in Section \ref{CH4:theory}.}
    \label{fig:haloprofile}
\end{figure}

The observed surface brightness profiles of the radio halo at the three different frequency bands show very similar behaviour as a function of radius, as indicated in Figure\,\ref{fig:haloprofile} where the normalised profiles are shown for comparison. These profiles are flatter than expected from models that assume a constant CRp density (or a declining CRp density, with positive values of $a$). Such a tendency was also observed in other radio halos such as the Coma Cluster \citep{Brunetti2002}. 
Assuming a value of $a=-0.5$ approximately reproduces the flatness of the observed profile as a function of radius, so we set this as a reference value in the following calculations. 

\begin{figure}[tbh]
    \centering
    \includegraphics[width=1.0\columnwidth]{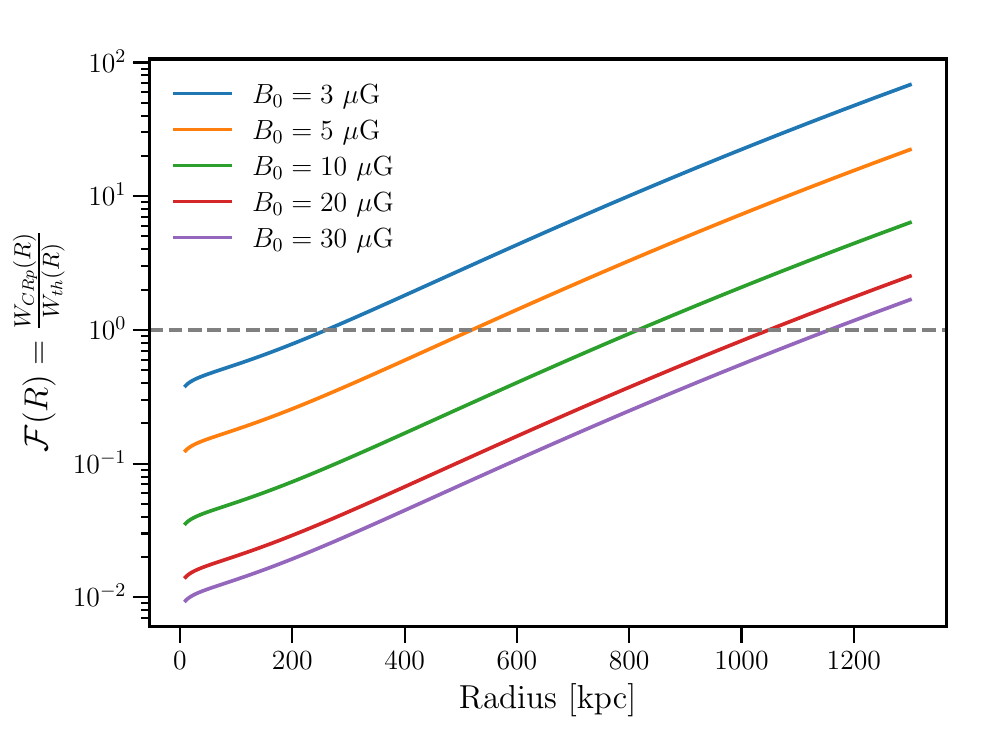}
    \caption{Required fractional energy density of cosmic ray protons with respect to the thermal protons as a function of radius to match the radio observations of the radio halo in Abell 2256 with hadronic models that have different central magnetic field strengths.}
    \label{fig:XCRPradius}
\end{figure}

To match the total synchrotron luminosity of the radio halo for $B_0=[3, 5, 10, 20, 30] \, \mu$G, hadronic models require an energy budget of CRp that is equal to {$[15, 4.9, 1.4, 0.6,0.4]$} times the thermal energy density averaged over the cluster volume within $R_\mathrm{500}$ respectively. This energy budget is large, because of the combination of the flat radio profile and steep synchrotron spectrum and improbable given the fact that the integrated CRp energy density is expected to be on the order of a few per cent of the total energy density in clusters \citep{Pinzke2010}. 
Figure\,\ref{fig:XCRPradius} shows that for all models, the radial profile of the cosmic ray energy density would exceed the thermal energy density within $R_\mathrm{500}$.\footnote{{We note that assuming $p_{p,\mathrm{min}}=0.01 mc$ would imply a CR energy budget that is about 40 percent higher than that in Fig. \ref{fig:XCRPradius}.}} Such energy budgets of CRp should result in a detectable gamma-ray luminosity and flux. 

\begin{figure}[thb]
    \centering
    \includegraphics[width=1.0\columnwidth]{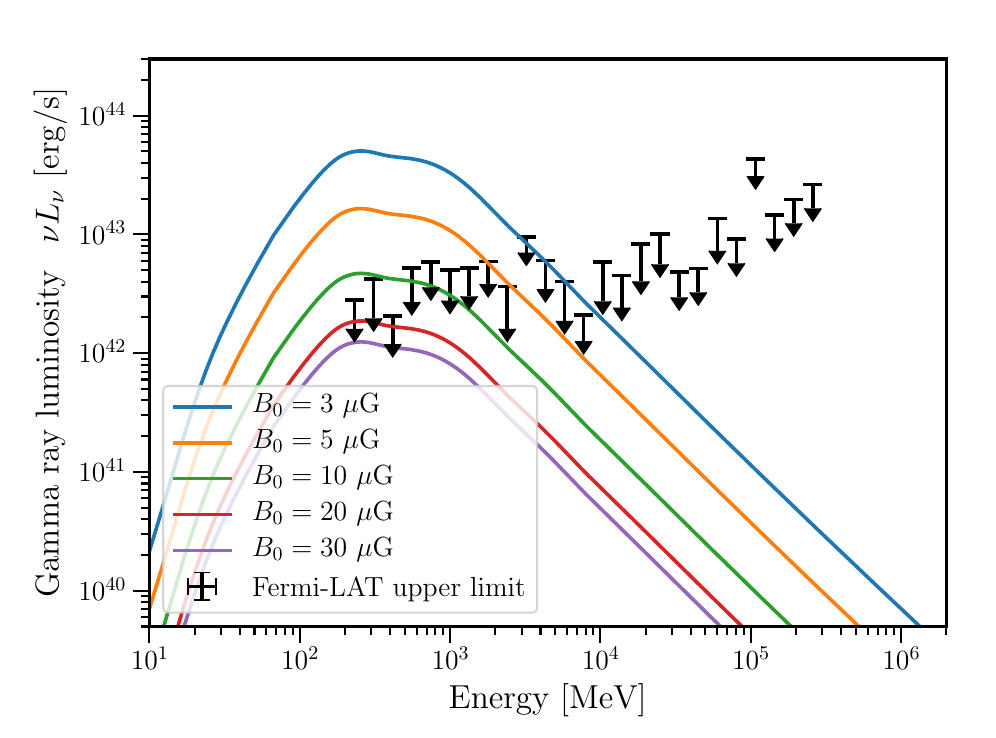}
    \caption{Expected gamma-ray luminosity from hadronic models that match the observed radio luminosity and brightness profile of the radio halo. Upper limits on the gamma-ray luminosity in Abell 2256 from Fermi-LAT observations are shown in black at 95\% confidence.}
    \label{fig:gammaray}
\end{figure}

To calculate the integrated synchrotron luminosity and gamma-ray luminosity from the hadronic model, we integrated out to $R_\mathrm{500}=1273$ kpc, although this cutoff is not sharp in practice. Therefore, this results in a conservative estimate for the expected gamma-ray radiation from the hadronic model. 
We also note that for $a=-0.5$ although the cosmic ray fraction increases away from the cluster centre, the gamma-ray luminosity still declines as a function of radius for $a>-1$. We show the expected gamma-ray flux derived from purely hadronic models that match the radio observations in Figure\,\ref{fig:gammaray}, where the overlay shows the current observational limits from Fermi-LAT. It is clear that for typical magnetic field values of $B_0=1$--$10\,\mu$G, gamma-rays would be detected if the halo was purely hadronic. At a three-sigma confidence level, the purely hadronic model disagrees with $B_0<17\, \mu$G. 

\section{Discussion}\label{CH4:discussion}
\begin{table*}[htb]
\centering
\caption{Radio halos detected over a large frequency range.}
\label{tab:OtherHalos}
\resizebox{2\columnwidth}{!}{%
\begin{tabular}{@{}lllllll@{}}
\toprule
Name &
  \begin{tabular}[c]{@{}l@{}}Frequency\\ {[MHz]}\end{tabular} &
  Spectral index\tablefootmark{a} &
  Curvature &
  $\sigma_\mathrm{2D}$\tablefootmark{b} &
  \begin{tabular}[c]{@{}l@{}}Mass\\ {[10$^{14} M_\odot$]}\end{tabular} &
  Reference \\ \midrule
Abell 2256 &
  23-1500 &
  $\alpha_\mathrm{23}^\mathrm{1500} = -1.56 \pm 0.02$ &
  no &
  $\approx 0.2$ &
  $6.2 \pm 0.1$ &
  This work \\
Bullet cluster &
  1100-3100 &
  $\alpha_\mathrm{1100}^\mathrm{3100} = -1.1 \pm 0.2$ &
  no &
  - &
  $13.1 \pm 0.29$ &
  \begin{tabular}[c]{@{}l@{}}\citep{Shimwell2014,Sikhosana2023}\end{tabular} \\
Toothbrush cluster &
  147-4900 &
  $\alpha_\mathrm{147}^\mathrm{4900} = -1.15 \pm 0.06$ &
  no &
  $<0.04$ &
  $10.8 \pm 0.45$ &
  \citep{Weeren2012} \\
Abell 2744 &
  325-1500 &
  $\alpha_\mathrm{325}^\mathrm{1500} = -1.32 \pm 0.14$ &
  no & 
  - &
  $9.8 \pm 0.4$ &
  \citep{Pearce2017} \\
Abell S1063 &
  325-3000 &
  $\alpha_\mathrm{325}^\mathrm{1500} = -0.94 \pm 0.08$ &
  yes &
  - &
  $11.4 \pm 0.34$ &
  \citep{Xie2020} \\
Coma cluster &
  30-5000 &
  $\alpha_\mathrm{144}^\mathrm{342} = -1.0 \pm 0.2$ &
  yes &
  - &
  $7.2 \pm 0.1$ &
  \citep{Bonafede2022} \\
MACS J0717.5+3745 &
  144-1500 &
  $\alpha_\mathrm{144}^\mathrm{1500} = -1.39 \pm 0.04$ &
  yes &
  $\approx 0.3$ &
  $11.5 \pm 0.5$ &
  \citep{Rajpurohit2021} \\ \bottomrule
\end{tabular}%
}
\tablefoot{\tablefoottext{a}{In case the spectrum is curved, only the spectral index measured below the break frequency, is given.}\tablefoottext{b}{Observed spatial scatter in the spectral index.}}
\end{table*}

The radio halo in Abell 2256 was among the first radio halos to be discovered \citep{Bridle1976}, with deeper follow-up data uncovering its progressively larger extent \citep[e.g.][]{Clarke2006,Brentjens2008,Owen2014}. The most recent estimate from \citet{Rajpurohit2023} shows that the largest linear size of the radio halo is at least 900 kpc. 

In this work, we find the radio halo to be significantly larger than these previous estimates, with an observed size at 144 MHz of 1.6 Mpc. This significant increase in observed size can be attributed to the unparalleled sensitivity of LOFAR at low frequencies, particularly to large-scale emission because of the many short baselines. The large size of the radio halo implies that a large fraction of the cluster volume is occupied by relativistic electrons and magnetic fields, which is in line with recent works that have also found that radio halos extend out to large radii when observed with high sensitivities at low frequencies \citep{Shweta2020,Cuciti2022,Botteon2022SciA}. In fact, it is likely that the observed size of the radio halo is still limited by missing short baselines, as data imaged without 100$\lambda$ baseline cuts shows a significantly larger and brighter radio halo (approximately 20\%) than data imaged without short \textit{uv} spacings. This was also found in previous works by injection of large mock radio halos into LOFAR data \citep{Bruno2023}. With the anticipated LOFAR2.0 upgrade to the LBA system, which can probe larger angular scales than the HBA system, observations will become more sensitive allowing the detection of even larger scale emission in nearby clusters. 

\subsection{Spectral properties of the halo}
The integrated spectrum of the radio halo in Abell 2256 is classified as ultra-steep and shows no indication of curvature (Figs. \ref{fig:intspixhalo}, \ref{fig:spixradius}). It is one of the few radio halos that are detected over a large frequency range, with other examples being the Bullet cluster \citep{Liang2000,Shimwell2014,Sikhosana2023}, the Toothbrush cluster \citep{Weeren2012,Gasperin2020}, Abell 2744 \citep{Pearce2017}, Abell S1063 \citep{Xie2020}, Coma \citep{Bonafede2022} and MACS J0717.5+3745 \citep{Rajpurohit2021}. We compiled the properties of these clusters in Table \ref{tab:OtherHalos}. It is interesting that the first three of these other radio halos do not show any indication of spectral curvature, with relatively flat spectra $\alpha < -1.3$ up to GHz frequencies, while the last three halos do show spectral curvature, resulting in ultra-steep spectra ($\alpha<-1.5$), at frequencies above $\sim1$ GHz. Abell 2256 thus presents a unique radio halo with an ultra-steep spectrum up to GHz frequencies, without spectral curvature.

Simple homogeneous turbulent re-acceleration models, with constant magnetic field and acceleration rate throughout the volume, have been successful in reproducing the observed counts and redshift distribution of radio halos in statistical samples \citep{Cassano2023}. In such models, ultra-steep spectra are expected above a cut-off frequency that scales with the acceleration efficiency in the ICM, which depends on the energetics (e.g. mass and mass ratio) of the merger \citep{Cassano2005,Cassano2006}.
It is therefore interesting that the integrated spectrum of the radio halo in Abell 2256 shows no curvature, while it is ultra-steep.

Variations in the magnetic field, turbulent energy and resulting acceleration efficiency throughout the emitting volume may complicate the apparent spectral behaviour. The superimposition of different regions can stretch the spectrum and generate a quasi-power-law spectrum when integrated over the full halo region. This effect has been observed in simulations \citep[e.g.][]{Donnert2013}, although they are limited in resolution and do not capture the full complexity of the dynamics of the ICM and CRs. The observed significant curvature and spectral index variations across the radio halo volume (e.g. Fig. \ref{fig:curv}), which were also observed at higher frequencies \citep{Rajpurohit2023}, point to an inhomogeneous situation in the Abell 2256 halo volume. In such a scenario, the steep spectral slope measured for Abell 2256 implies that a significant fraction of the emission in the halo volume is generated at low frequencies, where the acceleration time is shorter than the cooling time.

The intrinsic 2D scatter of the spectral index can be estimated as $\sigma_\mathrm{2D}=\sqrt{\sigma^2_\mathrm{obs} - \sigma^2_\mathrm{rms}}$, where $\sigma^2_\mathrm{obs}$ is the total observed scatter, and $\sigma^2_\mathrm{rms}$ is the scatter expected from the flux density uncertainties. Values of $\sigma_\mathrm{2D}$ of $0.14$ and $0.24$ are obtained for the scatter measured between $46-144$ and $23-46$ MHz, respectively. These variations are found to be quite large with respect to the other non-curved radio halo in the Toothbrush cluster \citep[$\sigma_\mathrm{2D}<0.04$;][]{Weeren2016}, as listed in Table \ref{tab:OtherHalos}.
However, the variations are of the same magnitude and spatial scale as those observed in MACS J0717.5+3745 \citep[$\sigma_\mathrm{2D}\approx 0.3$][]{Rajpurohit2021},
where an inhomogeneous situation was also proposed. 
Furthermore, Table \ref{tab:OtherHalos} indicates that MACS J0717.5+3745 has the steepest spectrum below the break frequency, implying that the level of inhomogeneity might be correlated with the steepness of the radio spectrum. It is also noteworthy that Abell 2256 is the least massive galaxy cluster in this sample, which implies that it has a smaller turbulent energy budget and will preferentially emit lower frequency radiation. However, this sample of clusters with radio halos detected over a large frequency range is small and the selection is not unbiased, thus additional data are required to draw definite conclusions. 

An inhomogeneous turbulent scenario has also been explored in the case of radio bridges, where theoretical models based on second-order Fermi re-acceleration predict that the fraction of the synchrotron emitting volume increases at lower frequencies \citep{BrunettiVazza2020}. The spectrum of radio bridges is not well known over large frequency ranges due to their low surface brightness, but the conditions for generating synchrotron emission in the volume (i.e. the acceleration time is smaller or equal to the cooling time) are more likely to be matched at lower emitting frequencies. However, it is still an open question how this process would result in a straight power-law for the integrated spectrum. Thus, explaining the combination of inhomogeneity in the halo volume and the perfect integrated power-law over multiple orders of magnitude in frequency as observed in Abell 2256 requires further theoretical studies.

\subsection{Testing a hadronic origin}
The radio halo in Abell 2256 is the nearest one in the universe that shows an ultra-steep spectrum below GHz frequencies. It is therefore one of the best candidates to put constraints on hadronic models from the combination of gamma-ray and radio data, as such a steep spectrum requires a large energy budget of cosmic ray protons which should result in observable gamma-ray emission. In Section \ref{CH4:gammaray}, we have shown that secondary models may explain the levels of radio and gamma-ray emission in Abell 2256 only in the case that {$B_0>17\mu$G}. This is significantly higher than typical magnetic field values of $B_0<10\mu$G estimated from Faraday rotation measurements in clusters \citep[e.g.][]{Bonafede2010,Vacca2012,Govoni2017,Osinga2022}. In fact, such strong magnetic fields are also unlikely for energetic reasons, since it would imply a magnetic pressure in the ICM that is $\geq$ 19\% of the thermal pressure, and a total non-thermal pressure (i.e. magnetic + CR) of the same order as the thermal pressure at $r=R_\mathrm{500}$.
This is significantly higher than the non-thermal pressure found observationally ($\sim$6\%) from the combination of X-ray and SZ observations \citep{Eckert2019}. Thus, in practice, assuming a hadronic origin of the halo, the combination of our LOFAR and gamma-ray data requires an untenable energy budget due to the combination of steep spectrum and flat radio brightness profile of the radio halo. We conclude that the purely hadronic model cannot explain the radio halo in Abell 2256.

This conclusion is quite robust, because of the conservative assumptions made in Section \ref{CH4:gammaray}. Firstly, we limit the integration of the gamma-ray emission at $r=R_\mathrm{500}$. The required energy budget for the non-thermal components would be even larger with a larger aperture radius. Secondly, similar to the case of the Coma cluster \citep[e.g.][]{Brunetti2012,Brunetti2017}, a flatter profile of the magnetic field would help reduce the energy budget of cosmic ray protons, but would not solve the tension because the magnetic field in the outskirts would become a dominant source of pressure. 

Additionally, we note that our models for the gamma-ray emission in Abell 2256 did not consider other possible sources of cosmic-ray protons. Shock (re)accelerated electrons that generate the bright radio shock in Abell 2256 may also generate gamma-rays via inverse Compton scattering off the CMB, provided that TeV electrons are accelerated at the shock. Additionally, protons should also be accelerated by the shock front, but the acceleration efficiency of cosmic ray protons at ICM shocks is poorly constrained \citep[e.g.][]{Vazza2015}, making it difficult to include this in our models. In any case, this implies that the central magnetic field strength would need to be even higher than $B_0=17\mu$G to explain the non-detection of gamma rays, which we have argued cannot be the case due to energetic reasons. In fact, our models are also conservative due to the fact that the radio halo was significantly  brighter (factor 2) in images without a $100\lambda$ \textit{uv} cut, but we employed this cut to make a fair comparison between different frequencies. We note, however, that  gamma-ray observations do not suffer from resolving out large-scale emission like radio observations do. 

In the turbulent re-acceleration scenario, a mildly relativistic `seed' population of electrons is re-accelerated by turbulence, which produces the radio halo \citep{BrunettiBlasi2005,BrunettiLazarian2011,Pinzke2017,Nishiwaki2022}. The origin of the seed electrons is still unconstrained, and hadronic interactions might produce the seeds for re-acceleration \citep[e.g.][]{Nishiwaki2022}. Jointly modelling the seed population from hadronic interactions at a level consistent with the upper limits presented here and the re-acceleration of those seed particles through turbulent magneto-hydrodynamics can address this problem, although such modelling is beyond the scope of the current paper. We can however make a qualitative assessment of this model. In the turbulent scenario, the emission is generated with a ratio of radio to gamma-rays that is typically a factor $3-10$ smaller than that in the case of purely hadronic models, thus allowing an energy budget of CRps that is up to one order of magnitude smaller than in the purely hadronic case. Current gamma-ray limits constrain the energy budget of the CRp (and magnetic field) to a level that is several times smaller than that obtained in Section \ref{CH4:gammaray}. If the radio halo is indeed generated by turbulent re-acceleration, \citet{Brunetti2009} predicted that a gamma-ray detection would only be possible in the case that $B_0<1\,\mu$G. The non-detection is thus consistent with typical magnetic field strengths between $1-10\,\mu$G that are observed in clusters from Faraday rotation experiments \citep[e.g.][]{Osinga2022}. The current Fermi-LAT limits do not rule out re-acceleration of secondary particles for the origin of the halo in Abell 2256, as was also concluded for the Coma cluster \citep{Brunetti2017,Adam2021}. 

\subsection{Diffusive shock acceleration in the radio shock}
The diffusive shock acceleration (DSA) of fossil electrons is the most promising model for radio shocks in clusters \citep[e.g.][]{Pinzke2013,Vazza2015,Vazza2016,KangRyu2016}. According to DSA, the integrated spectral index of radio shocks cannot be flatter than $\alpha=-1.0$. However, this constraint was violated in the radio shock in Abell 2256 with early LOFAR observations at low frequency, where a radio shock spectral index value of -0.85$\pm$0.01 was found \citep{Weeren2012}. {Observations between 100 MHz and 3 GHz show no such violation, with a recent study by \citet{Rajpurohit2022b} finding a spectral index of $\alpha=-1.07 \pm 0.02$ between 144 MHz and 3 GHz.}

In this work, we found that the low-frequency spectral index of the relic is $\alpha_{23}^{146}=-0.87\pm0.06$, {which is in line with the previous study by \citet{Weeren2012}, and shows a discrepancy with DSA. However, when combining our data with higher frequency data up to $3$ GHz, we obtain a value of $\alpha_{23}^{3000}=-1.00\pm0.02$, which is consistent with DSA, and slightly flatter than the value found by \citet{Rajpurohit2022b}.} {The flatter spectral index observed between 23 and 144 MHz could be caused by flux scale biases, which are more impactful when the frequency difference of the associated flux measurements is small. As \citet{Rajpurohit2022b} noted, the original LOFAR HBA calibration provided fluxes that were too high, but this issue was addressed by re-scaling the flux scale using compact bright sources and we have taken over the scaling in this study. However, if the HBA flux scale is still too bright, this would cause \citet{Rajpurohit2022b} to overestimate the steepness of the spectral index above 144 MHz and this study to underestimate the steepness of the spectral index below 144 MHz. Given the possible systematic issues with the LOFAR HBA flux scale, we only draw conclusions from the spectrum evaluated over the entire range of available frequencies, and we see no strong evidence of the spectrum between 24 and 3000 MHz being inconsistent with DSA.}

Similar to other radio shocks that are mapped over wide frequency ranges, such as the Toothbrush and Sausage radio shock \citep{Rajpurohit2020,Loi2020}, our findings suggest that there is no deviation from a power-law over multiple orders of magnitude, indicating no inconsistency with the standard DSA scenario in Abell 2256. The radio shock interpretation is also consistent with the X-ray detection of a nearby shock in Abell 2256 by \citet{Ge2020}.
However, while the DSA interpretation seems to be supported by observations, some problems remain to be understood. In the case of standard DSA, an integrated spectrum with a spectral index close to $\alpha=-1$ requires a large Mach number \citep{Blandford1987}, which is inconsistent with what has been measured for the Mach number of the X-ray detected shock in Abell 2256 \citep[SF1 in][]{Ge2020}. 
This might be resolved by considering that the radio shock region consists of an ensemble of shocks, and the radio and X-ray observations trace different parts of this distribution, with projection effects also playing a significant role \citep[e.g.][]{Wittor2021}. 

\subsection{Origin of AGN related sources}
The physical interpretation and age estimation of the various smaller ultra-steep spectrum sources in Abell 2256 have been complicated by the inability of previous studies to fit their spectra with simple synchrotron models, due to the strong curvature implying low break frequencies \citep[e.g.][]{Brentjens2008,Weeren2012Abell,Owen2014}. The new ultra-low frequency data show that we can now observe many of the radio source spectra flatten towards lower frequencies (Fig. \ref{fig:AGNspectra}).

The question of whether the F-complex should be considered a radio shock was raised by \citet{Owen2014}, because of its steep spectrum, polarisation and elongated structure. However, unlike the large radio shock of Abell 2256, the spectrum of this source is strongly curved, resembling a typical aged AGN spectrum. As raised already by \citet{Bridle1979}, sources F1 and F2 might all be part of the tail of source F3. 
We propose that sources F2 and F1 are related to the Fabricant Galaxy 122 (FG122) at the location of F3. The synchrotron modelling implied that the radiative age of the sources is approximately 200 Myr, which is consistent with the time it would take FG122 to travel the distance between its current location and the location of the F complex, given the typical velocity dispersion in the cluster \citep{Brentjens2008}. If the magnetic field strength is lower than our assumed $7 \mu$G \citep{Brentjens2008}, then the age estimates would increase further and this picture would remain consistent with observations, unless the magnetic field is significantly weaker than $B=1.8\mu$G, in which case inverse Compton losses would quickly dominate. Furthermore, we observe no spectral index gradient across sources F1 to F3 in the low-frequency spectral index map (Fig. \ref{fig:radiofullspix}), which is expected in the standard spectral ageing scenario \citep[e.g.][]{Myers1985}, for a constant magnetic field when observing sources below the break frequency.

Interestingly, a new source was detected below the F complex which complicates the scenario once more. We have named this source F4. The spectrum of F4 remains curved below 100 MHz, with a spectral index of $\alpha_{23}^{144}=-1.9 \pm 0.1$, indicating that we have not yet found the break frequency of this source, but constrain it to be $<23$ MHz. Whether the source is physically related to the F1-F3 complex is difficult to say. However, the sudden steepening in the spatial spectrum, with no gradient in the spectral index map between F2 and F4 makes a physical relation unlikely. Multiple cluster members are located in the region co-spatial with F4, so an optical association is difficult to make correctly, given the diffuse morphology of the source. However, the morphology of the radio emission and the spectral index map shown in Figure \ref{fig:opticalFcomplex} indicate a possible host galaxy (MCG+13-12-020). Given the steep spectrum of the source at such low frequencies, it is likely that source F4 is a very old remnant radio galaxy with an age of $>400$ Myr. The dense and turbulent {intracluster} medium possibly quenched the expansion of source F4, limiting adiabatic losses and allowing the low-frequency detection of such an old source \citep{Murgia2011}.

The source AG+AH is located at approximately 800 kpc from the head of the tailed source C and shows a curved spectrum where $\alpha_{144}^{351}=-2.05$ \citep{Weeren2012Abell} while we observe $\alpha_{23}^{144}=-0.91\pm0.07$. In previous LOFAR observations, \citet{Weeren2012Abell} noted that if the break frequency of the spectrum is below 50 MHz, the radiative age of the source would be old enough to link it to source C. However, we observed the break frequency at $113\pm12$ MHz, implying AG+AH can only be related if the fossil plasma is re-accelerated. Processes such as the gentle re-energisation process \citep[e.g.][]{Gasperin2017} or a shock wave that is also responsible for the radio shock can increase the age of the source substantially, \citep[e.g.][]{Kale2012}, allowing a physical relation between the sources. Such processes could also explain the filamentary `ribs' coming off the radio source, which are likely caused by complex interactions of the fossil plasma with the environment \citep{Rudnick2021}. Interestingly, like the first ribbed source detected in Abell 3266, AG+AH is also related to an apparently one-sided tail. There are multiple sources now found in clusters that show such one-sided tails with such rib-like features, including IC1711 in Abell 1314 \citep{Wilber2018}, and SDSS J105851.01+564308.5 in Abell 1132 \citep{Wilber2019}. These observations may provide insights into the origin of these phenomena. 

Finally, source AI was discovered by \citet{Weeren2009}, where it was suggested to be either a radio shock or a radio phoenix. It was recently classified as a radio phoenix based on the morphology, location and curved spectrum by \citep{Rajpurohit2023}. This is corroborated by the ultra-low frequency results here, where the spectrum indeed approaches a typical AGN spectrum with $\alpha_{23}^{46}=-1.18$ which significantly steepens towards higher frequencies. 

\section{Conclusion}\label{CH4:conclusion}
We have investigated particle acceleration in Abell 2256 by studying the lowest energy electrons observable by ground-based telescopes. This study presented the first high-quality LOFAR observations down to 16 MHz of Abell 2256, proving the potential for new cluster science with LOFAR ultra-low frequency observations. The radio halo, radio shock, and most prominent fossil plasma sources in Abell 2256 were all detected clearly at 144, 46 and 23 MHz. The ultra-low frequency data paint a consistent picture with respect to what was found at higher frequencies, where both the radio halo and radio shock show straight power-law spectra over multiple orders of frequency, while the fossil plasma sources show relatively flat spectra at low frequencies that can curve extremely towards higher frequencies. This dichotomy, where spectral shapes are powerfully distinguished, that starts to show at low frequencies could help in the classification of diffuse cluster sources, which is becoming increasingly challenging as cluster radio emission is more ubiquitously detected.

We summarise the main results of this work as follows:
\begin{enumerate}
    \item The combination of low-frequency radio and gamma-ray data places some of the strongest direct constraints on the purely hadronic model for radio halos. The data are only consistent with the purely hadronic model for central magnetic field strengths $>17\,\mu$G, which are improbably high given  non-thermal pressure and magnetic field constraints that exist for comparable clusters. 
    This is only the second cluster for which such a direct constraint was produced, with the only other cluster being the Coma cluster, where data also disfavours a purely hadronic model.

    \item The sensitive LOFAR HBA image shows that the radio halo has a largest linear size of 24 arcminutes at 144 MHz, corresponding to a linear size of 1.6 Mpc at the cluster redshift. This is larger than previously measured.
    
    \item The integrated radio halo spectrum follows a straight power-law with a spectral index of $-1.56\pm0.02$ over a wide frequency range from 24 to 1500 MHz. The core region emits flatter spectrum emission ($\alpha=-1.36$) than the overall radio halo, and the wedge arc between the radio shock and the F-complex shows somewhat steeper emission. 
    
    \item Although the integrated spectrum follows a straight power-law, we found significant spatial variations in the spectral index and curvature across the radio halo on the order of $\sigma(\alpha_\mathrm{2D})=0.2$. This implies that the emitting volume is strongly inhomogeneous, which is difficult to reconcile with the perfect power-law of the integrated spectrum by current theories.
    
    \item The radio shock spectrum also agrees with a straight power-law, but is significantly flatter than the radio halo, with $\alpha=-1.00\pm0.02$ between 24 and 3000 MHz. The spectral index map at low frequencies also shows steepening from the southwest side to the northeast side, indicating the direction of the shock as electrons age in the downstream region. 
    
    \item Abell 2256 hosts six complex radio sources with mostly curved spectra, of which five were known previously. We have detected a new ultra-steep spectrum source just below the F-complex, which we have named F4. While we see the spectra of the other complex radio sources flatten significantly towards 23 MHz, F4 still shows an ultra-steep spectral index of $\alpha_{23}^{144}=-1.9 \pm 0.1$, and we suspect it is unrelated to sources F1-F3 based on the sudden discontinuity in the spectral index map.
    
    \item We have modelled the synchrotron emission of these complex radio sources, finding typically curved spectra that agree well with simple ageing models, and finding radiative ages around 200 Myr. These findings are consistent with the interpretation that these are fossil plasma sources. 
    
\end{enumerate}
   
Most of the understanding about the origin and formation of diffuse radio emission in clusters has been derived from studies of relatively massive galaxy clusters that could be detected at GHz frequencies. However, turbulent re-acceleration models predict that an increasing fraction of halos in lower mass clusters should have a steep spectrum \citep[e.g.][]{Cassano2010}, implying they are missed at high frequencies. To constrain model parameters, a large lever arm is needed for precise spectral index determination. Observations down to about 16 MHz can provide a similar lever arm when combined with $\sim 150$ MHz to the lever arm historically used by combining 150 and 1500 MHz observations. The successful observations made in the lowest radio window available to ground-based telescopes thus open up exciting possibilities for future research on particle acceleration mechanisms in clusters.

\begin{acknowledgements}
      EO and RJvW acknowledge support from the VIDI research programme with project number 639.042.729, which is financed by the Netherlands Organisation for Scientific Research (NWO). 
  FdG acknowledges the support of the ERC CoG grant number 101086378.
      AB acknowledges financial support from the European Union - Next Generation EU.
LOFAR \citep{Haarlem2013} is the Low-Frequency Array designed and constructed by ASTRON. It has observing, data processing, and data storage facilities in several countries, which are owned by various parties (each with their own funding sources), and that are collectively operated by the ILT foundation under a joint scientific policy. The ILT resources have benefited from the following recent major funding sources: CNRS-INSU, Observatoire de Paris and Université d’Orléans, France; BMBF, MIWF-NRW MPG, Germany; Science Foundation Ireland (SFI), Department of Business, Enterprise and Innovation (DBEI), Ireland; NWO, The Netherlands; The Science and Technology Facilities Council, UK; Ministry of Science and Higher Education, Poland; The Istituto Nazionale di Astrofisica (INAF), Italy. This research made use of the LOFAR-UK computing facility located at the University of Hertfordshire and supported by STFC [ST/P000096/1], and of the LOFAR-IT computing infrastructure supported and operated by INAF, and by the Physics Dept. of Turin University (under the agreement with Consorzio Interuniversitario per la Fisica Spaziale) at the C3S Supercomputing Centre, Italy. 
    We thank the staff of the GMRT that made TGSS possible. GMRT is run by the National Centre for Radio Astrophysics of the Tata Institute of Fundamental Research.
EO thanks Jonah Wagenveld and Roland Timmerman for supplying useful Python scripts and the Istituto di Radioastronomia at CRN Bologna for the hospitality during the spring of 2022, where helpful discussions took place.     
      
\end{acknowledgements}

\bibliographystyle{aa}
\bibliography{firstbib.bib}

\newpage

\begin{appendix}
\section{Flux measurements and decametre sky field-of view}\label{appendixI}
\begin{figure}[htb]
    \centering
    \includegraphics[width=1.0\columnwidth]{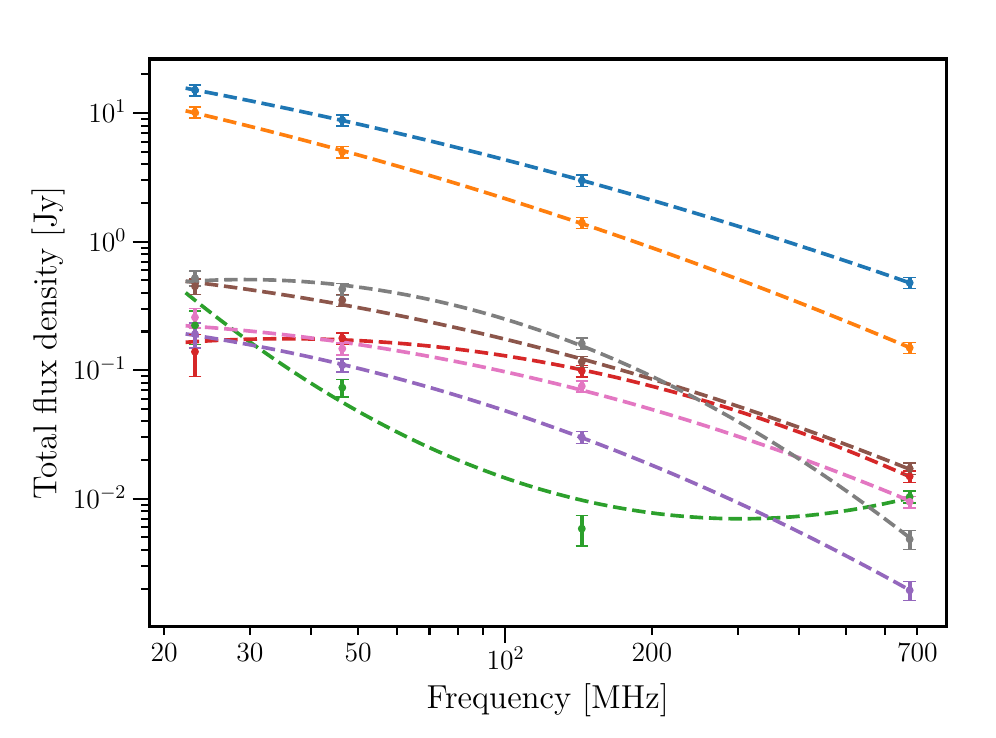}
    \caption{Flux density measurements at LOFAR LBA, HBA and uGMRT (675 MHz) frequencies with a best-fit polynomial in logspace shown. Most error bars are dominated by the assumed 10\% absolute flux density scale error.}
    \label{fig:fluxscale}
\end{figure}

To verify the flux density scale of the LOFAR LBA and HBA images in the direction of Abell 2256, we have compared our data with deep upgraded Giant Metrewave Radio Telescope (uGMRT) data at 675 MHz from \citet{Rajpurohit2022b}. We have identified eight compact bright sources around Abell 2256 which are visible in the LOFAR 24 MHz, 46 MHz, HBA and uGMRT images. The 24 MHz flux was calculated in a 90-arcsecond resolution map to make sure all flux was captured for point sources which may still suffer from residual ionospheric errors. We decided not to compare to ancillary VLA 1-4 GHz data, as the  field of view of those data is too small to make comparisons for many sources around Abell 2256. The results are shown in Figure\,\ref{fig:fluxscale}, where the HBA flux is corrected by a scaling factor of 0.83, and the LBA flux is not adjusted. {Most sources show a curved spectrum with flattening towards lower frequencies, which could either indicate a low flux density scale or a physical effect. We argue this is likely a physical effect, as it was also seen recently in the LoLSS survey \citep{Gasperin2023}, where most sources were found to have a curved spectrum between 54 MHz and GHz frequencies. There, it is clear that there is no significant flux scale issue, as the spectra were in line with observations at 38 MHz from the 8C survey. This indicates that at lower frequencies the spectrum physically flattens, an effect that we also observe in our ultra-low frequency observations.} Finally, the results show that the fluxes are in line with simple log-space polynomial fits, implying that there is no significant bias in the flux density scale. 

\begin{figure}[htb]
    \centering
    \includegraphics[width=1.0\columnwidth]{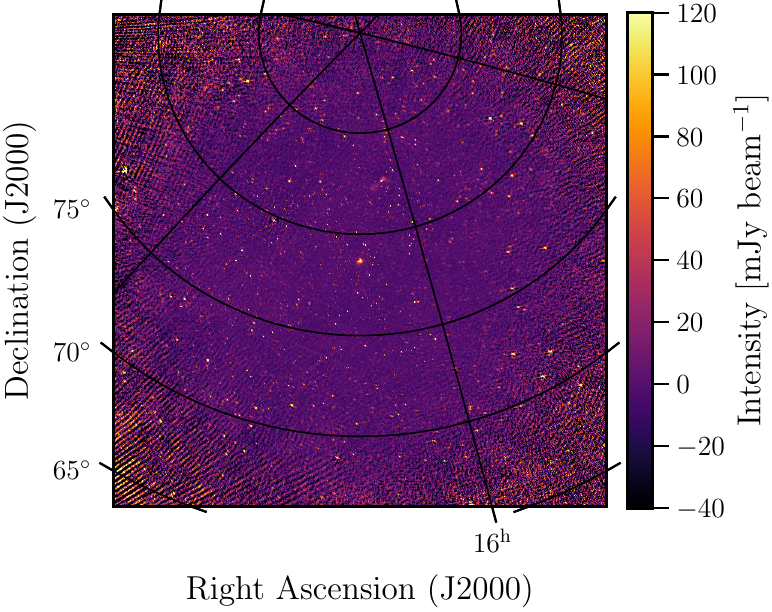}
    \caption{Full field-of-view image of the decametre sky from the LBA observations between 16-30 MHz. This image covers about 200 deg$^{2}$ and is centered on Abell 2256. The primary beam half-power beam-width is $\sim$9 degrees at 30 MHz and the restoring beam is 39$^{\prime\prime}\times24^{\prime\prime}$.}
    \label{fig:fullFOVLBA}
\end{figure}

For completeness, the flux measurements of the radio halo and radio shock regions defined in Sections \ref{CH4:radiohalo} and \ref{CH4:radioshock} are given in Table \ref{tab:halorelicflux}. \noindent The full field of view of the LBA observations in the 16-30 MHz range is shown in Figure\,\ref{fig:fullFOVLBA}.
\begin{table}[bht]
\centering
\caption{Flux density measurements and best-fit spectral index of the radio halo and radio shock regions as measured in Figures \ref{fig:intspixhalo} and \ref{fig:intspixrelic}.}
\label{tab:halorelicflux}
\resizebox{\columnwidth}{!}{%
\begin{tabular}{@{}lllll@{}}
\toprule
Source &
  \begin{tabular}[c]{@{}l@{}}$S_\mathrm{23MHz}$\\ {[Jy]}\end{tabular} &
  \begin{tabular}[c]{@{}l@{}}$S_\mathrm{46MHz}$\\ {[Jy]}\end{tabular} &
  \begin{tabular}[c]{@{}l@{}}$S_\mathrm{144MHz}$\\ {[Jy]}\end{tabular} &
  $\alpha$ \\ \midrule
Halo        & 12.72$\pm$1.38 & 4.79$\pm$0.52  & 0.91$\pm$0.11 & -1.44$\pm$0.08 \\
Wedge arc   & 2.69$\pm$0.27  & 0.94$\pm$0.10  & 0.15$\pm$0.02 & -1.57$\pm$0.08 \\
Halo core\tablefootmark{a}   & 1.70$\pm$0.17  & 0.66$\pm$0.07  & 0.14$\pm$0.01 & -1.36$\pm$0.08 \\
Halo total  & 15.41$\pm$1.40 & 5.73$\pm$0.53  & 1.06$\pm$0.11 & -1.46$\pm$0.07 \\
Shock R1    & 2.00$\pm$0.21  & 0.92$\pm$0.10  & 0.28$\pm$0.03 & -1.08$\pm$0.08 \\
Shock R2    & 5.75$\pm$0.58  & 3.18$\pm$0.32  & 1.25$\pm$0.13 & -0.83$\pm$0.08 \\
Shock R3    & 9.79$\pm$1.00  & 5.65$\pm$0.58  & 2.16$\pm$0.22 & -0.83$\pm$0.08 \\
Shock total & 18.73$\pm$1.35 & 10.19$\pm$0.74 & 3.78$\pm$0.28 & -0.87$\pm$0.05 \\ \bottomrule
\end{tabular}%
}
\tablefoot{\tablefoottext{a}{We note that the `halo core' region is a subset of the `halo' region}}
\end{table}

\section{Uncertainty maps}
\begin{figure*}[thb]
    \centering
    \includegraphics[width=1.0\textwidth]{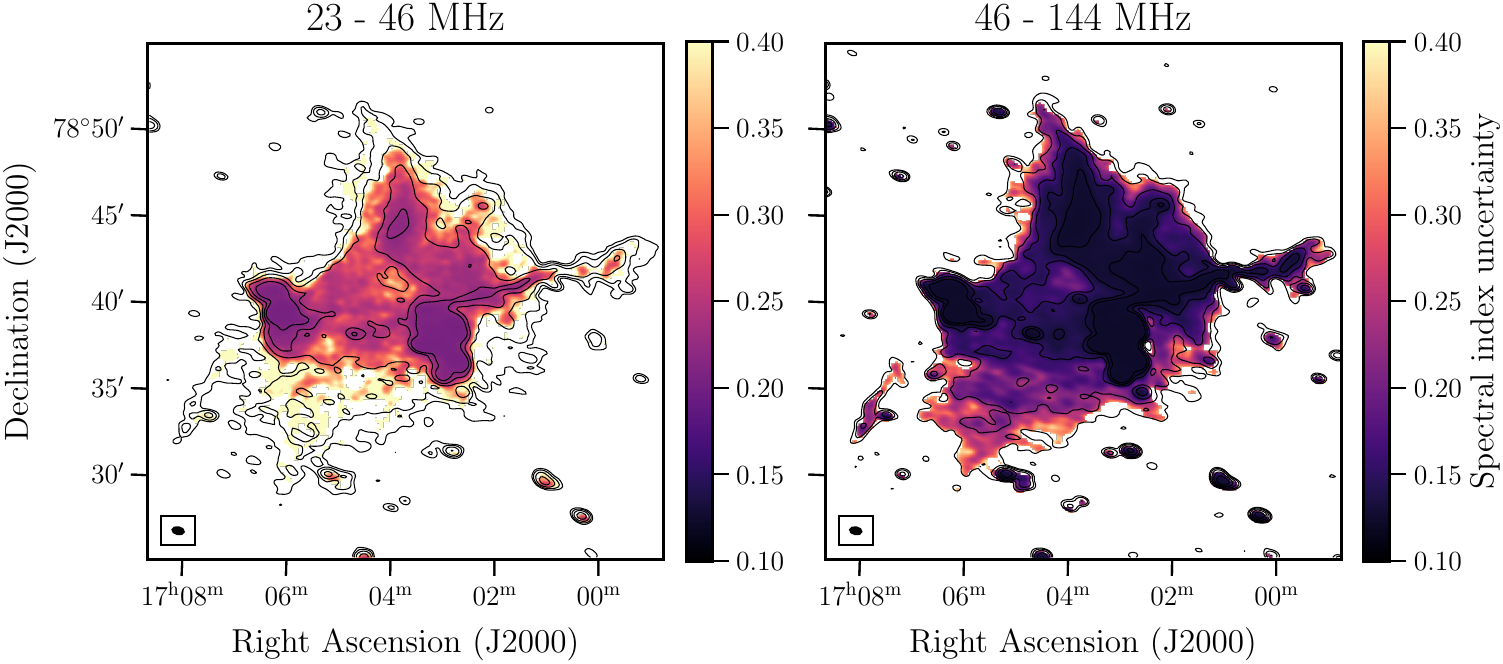}
    \caption{Spectral index uncertainty maps of Abell 2256, including a systematic 10\% flux scale uncertainty and map noise, at different frequencies with the restoring beam shown in the bottom left inset. Both maps have been smoothed to a common resolution of 39$^{\prime\prime}\times24^{\prime\prime}$. The median uncertainties are 0.31 and 0.19 in the left and right image, respectively. The contours show higher frequency [3, 6, 12, 24, 48]$\sigma$ levels where $\sigma$ denotes the background rms noise level.}
    \label{fig:spix_err}
\end{figure*}

\begin{figure}[htb]
    \centering
    \includegraphics[width=1.0\columnwidth]{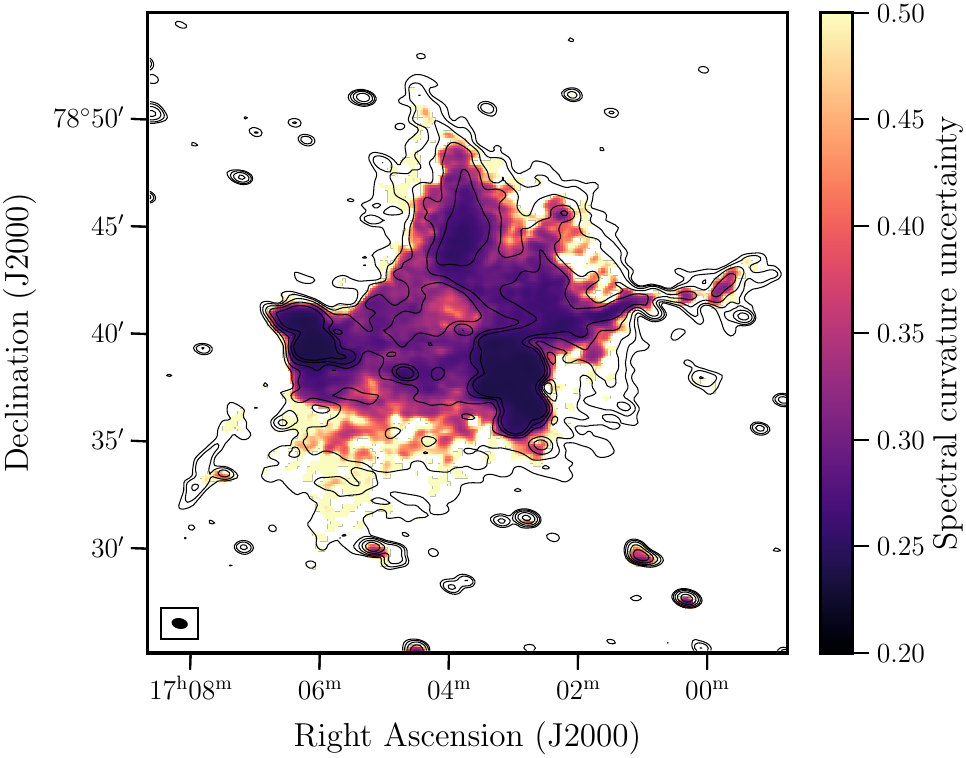}
    \caption{Curvature uncertainty map of Abell 2256, including a systematic 10\% flux scale uncertainty and map noise, with the restoring beam of 39$^{\prime\prime}\times24^{\prime\prime}$  shown in the bottom left inset. The median uncertainty is 0.33. The contours show 144 MHz frequency [3, 6, 12, 24, 48]$\sigma$ levels where $\sigma$ denotes the background rms noise level.}
    \label{fig:curv_err}
\end{figure}

We show in Figures \ref{fig:spix_err} and \ref{fig:curv_err} the uncertainty maps for the spectral index and spectral curvature respectively. The uncertainty on the spectral index was calculated as 
\begin{equation}\label{eq:spixunc}
    \Delta\alpha = \frac{1}{\ln(\nu_1/\nu_2)}\left[\left(\frac{\Delta S_1}{S_1}\right)^2 + \left(\frac{\Delta S_2}{S_2}\right)^2\right]
\end{equation}
where $\nu$ refers to the frequency of the observation, $S$ to the corresponding observed flux, and $\Delta S$ to the uncertainty on the flux ({which includes both the absolute flux scale uncertainty and the RMS map noise}). The uncertainty on the curvature map was computed from the uncertainties on the spectral index maps using standard error propagation. 

\section{Halo fitting}\label{CH4:halofit}
\begin{figure*}[htp]

\subfloat[144 MHz]{%
  \includegraphics[width=\textwidth]{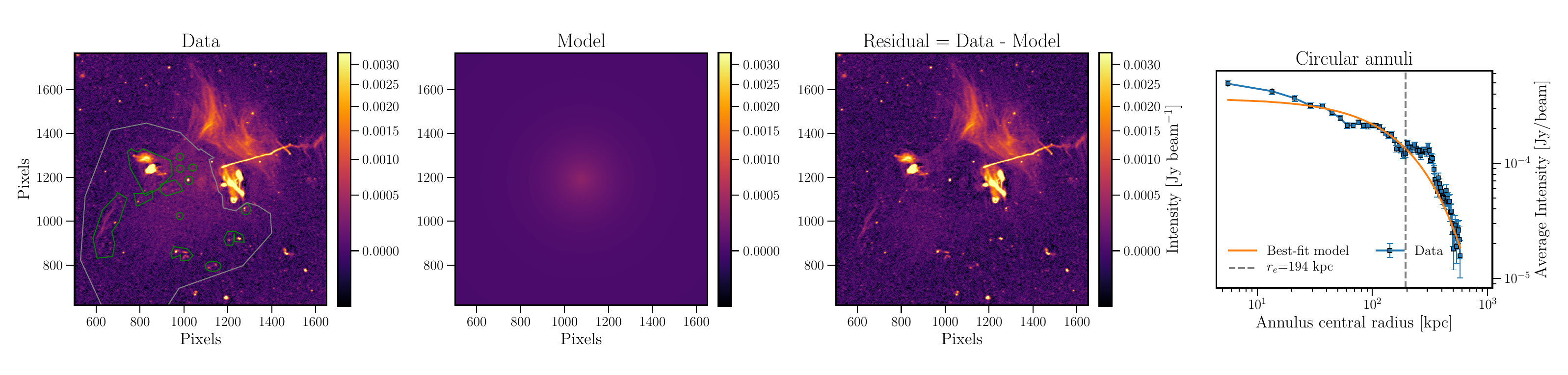}%
}

\subfloat[46 MHz]{%
  \includegraphics[width=1.0\textwidth]{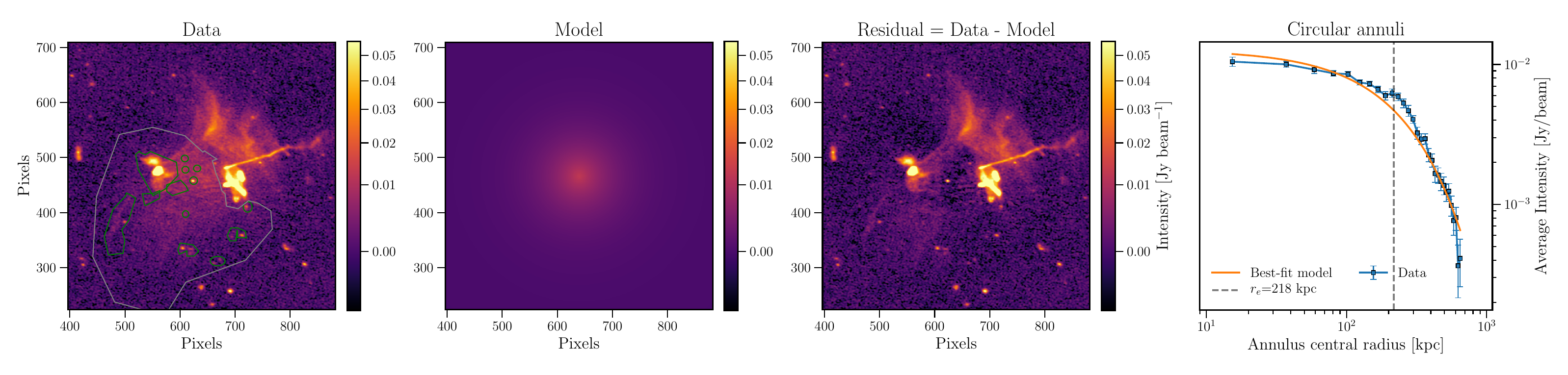}%
}

\subfloat[23 MHz]{%
  \includegraphics[width=1.0\textwidth]{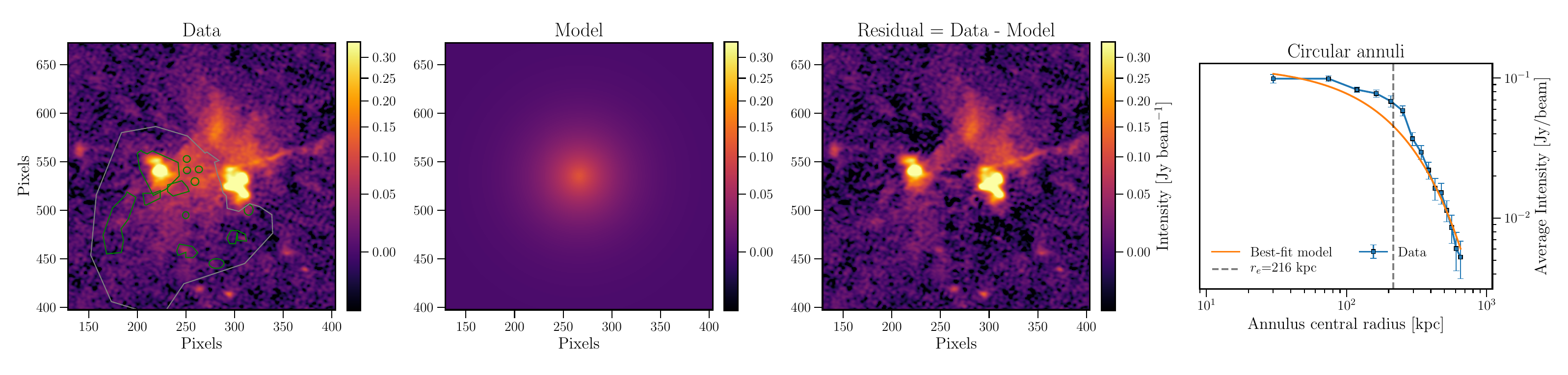}%
}

\caption{Halo fit results for various frequencies. The first panel shows the data, with masked sources indicated by the green regions and the fitting region indicated by the grey region. The second panel shows the best-fit halo model (Eq. \ref{eq:halomodel}) and the third panel the residual image. The last panel shows the same fit visualised in one dimension calculated from concentric annuli.} \label{fig:halofit}

\end{figure*}

Figure\,\ref{fig:halofit} shows the results of the Halo-Flux Density CAlculator \citep[Halo-FDCA;][]{Boxelaar2021}, a Markov-chain Monte Carlo code that fits a simple surface brightness model,
\begin{equation}\label{eq:halomodel}
I(r) = I_0 \exp(-r/r_e),
\end{equation}
to a radio halo. We have indicated the region used for the fitting, and the regions used to mask the compact AGNs in the leftmost panel. The resulting best-fit parameters are given in Table \ref{tab:halofit}.

\begin{table*}[p]
\centering
\caption{Results of the 2D halo-fitting. The uncertainties indicate statistical uncertainties only, computed from the 16th and 84th percentile of the Markov chain.}
\label{tab:halofit}
\begin{tabular}{@{}llllll@{}}
\toprule
\begin{tabular}[c]{@{}l@{}}Frequency\\ {[MHz]}\end{tabular} &
  \begin{tabular}[c]{@{}l@{}}$I_0$\\ {[Jy arcsec$^{-2}$]}\end{tabular} &
  \begin{tabular}[c]{@{}l@{}}RA\\ {[deg]}\end{tabular} &
  \begin{tabular}[c]{@{}l@{}}DEC\\ {[deg]}\end{tabular} &
  \begin{tabular}[c]{@{}l@{}}$r_e$\\ {[kpc]}\end{tabular} &
  $\chi^2_\mathrm{red}$ \\ \midrule
144 & $(9.878 \pm 0.007) \times 10^{-6}$ & $-103.9181\pm0.0002$ & $78.64251\pm0.00003$ & $193.6 \pm 0.1$ & 1.1 \\
46  & $(4.574 \pm 0.004) \times 10^{-5}$ & $-103.8822\pm0.0003$ & $78.64905\pm0.00004$ & $218.2 \pm 0.2$ & 1.5 \\
23  & $(1.130 \pm 0.001) \times 10^{-4}$ & $-103.8626\pm0.0002$ & $78.64970\pm0.00004$ & $216.1 \pm 0.2$ & 3.7 \\ \bottomrule
\end{tabular}%
\end{table*}

\end{appendix}

\end{document}